\documentclass[preprint,aps,tighten,epsfig,nofootinbib]{revtex4}
\usepackage{graphicx}

\begin{document}
\draft

\centerline{\bf Molecular Dynamics Simulation of Amyloid
$\beta$ Dimer Formation}

\smallskip

\noindent{\sc\scriptsize 
B.  Urbanc$^\ast$~\footnote{Corresponding author.  Email: 
brigita@bu.edu}, L.  Cruz$^\ast$, F.  Ding$^{\ast\dagger}$,
D.  Sammond$^\dagger$, S.  Khare$^\dagger$, S.  V. 
Buldyrev$^\ast$, H.  E.  Stanley$^\ast$, and N.  V. 
Dokholyan$^\dagger$}

\smallskip

\noindent{\scriptsize
$^\ast$Center for Polymer Studies, Department of Physics, Boston
University, Boston, MA 02215; \\
$^\dagger$Department of Biochemistry and Biophysics, University of
North Carolina at Chapel Hill, School of Medicine, Chapel Hill, NC 27599. \\
}

%======================================================================
\begin{center}
ABSTRACT
\end{center}
\noindent
{\bf Recent experiments with amyloid-$\beta$ (A$\beta$) peptide suggest
that formation of toxic oligomers may be an important contribution to
the onset of Alzheimer's disease.  The toxicity of A$\beta$ oligomers
depends on their structure, which is governed by assembly dynamics.  Due
to limitations of current experimental techniques, a detailed knowledge
of oligomer structure at the atomic level is missing.  We introduce a
molecular dynamics approach to study A$\beta$ dimer formation: (1) we
use discrete molecular dynamics simulations of a coarse--grained model
to identify a variety of dimer conformations, and (2) we employ all-atom
molecular mechanics simulations to estimate the thermodynamic stability
of all dimer conformations.  Our simulations of a coarse--grained
A$\beta$ peptide model predicts ten different planar $\beta$-strand
dimer conformations.  We then estimate the free energies of all dimer
conformations in all-atom molecular mechanics simulations with explicit
water.  We compare the free energies of A$\beta$(1--42) and
A$\beta$(1--40) dimers.  We find that (a) all dimer conformations have
higher free energies compared to their corresponding monomeric states,
and (b) the free energy difference between the A$\beta$(1--42) and the
analogous A$\beta$(1--40) dimer conformation is not significant.  Our
results suggest that A$\beta$ oligomerization is not accompanied by the
formation of stable planar $\beta$-strand A$\beta$ dimers.  \\ }

\smallskip
\noindent
{\bf Keywords:} Alzheimer's disease, coarse--grained peptide
model, monomer conformation, dimer conformation, free energy,
stability

\newpage
%======================================================================

%======================================================================
\newpage

\section{Introduction}

Alzheimer's disease (AD) is neuropathologically characterized by
progressive neuronal loss, extracellular amyloid plaques and
intracellular neurofibrillary tangles (Yankner, 1996; Selkoe, 1997).
Fibrillar amyloid plaques, a result of A$\beta$ peptide aggregation,
have been implicated in the pathogenesis of AD.  Recent experimental
studies on A$\beta$ peptide (Lambert et al., 1998; El-Agnaf et al.,
2000; El-Agnaf et al., 2001; Dahlgren et al., 2002) as well as
various animal model studies (Hsia et al., 1999; Mucke et al., 2000;
Dodart et al., 2002; Westerman et al., 2002; Walsh et al., 2002)
suggest that soluble forms of A$\beta$ assemblies cause substantial
neuronal dysfunction even before the appearance of amyloid plaques.
Hence, finding the conformation of these oligomeric forms of A$\beta$
may be important for understanding of neurotoxicity in AD (Kirkitadze et
al., 2002; Klein et al., 2001; Klein, 2002a; Klein, 2002b;
Bucciantini et al., 2002; Kayed et al., 2003).  At present, the
precise nature, conformation, and time evolution from monomer A$\beta$
peptides into intermediates is still unknown.

The fibrillar structure of A$\beta$ peptide aggregates is relatively
well established.  Experiments have targeted the structure of A$\beta$
fibrils using electron microscopy (Malinchik et al., 1998; Tjernberg
et al., 1999; Tjernberg et al., 2002), X-ray diffraction (Malinchik et
al., 1998; Serpell et al., 2000), electron paramagnetic resonance
spectroscopy (T\"or\"ok et al., 2002) and solid state nuclear magnetic
resonance spectroscopy (Balbach et al., 2002; Petkova et al., 2002;
Antzutkin et al., 2002; Thompson, 2003; Antzutkin et al., 2003).
The most common view is that A$\beta$(1-40) and A$\beta$(1-42) in
fibrils form parallel $\beta$-sheets with a $\beta$-turn between
residues Asp 23 and Lys 28.  The most flexible regions of the peptide in
a fibril are the first 10 amino acids of the N-terminus, last few amino
acids of the C-terminus (residues 39-42), and the $\beta$-turn region
between residues 23 and 28 (Petkova et al., 2002; T\"or\"ok et al.,
2002).

The aggregation process from a monomer A$\beta$ peptide via soluble
oligomeric states to fibrils is a complex dynamic event which depends
critically on the peptide concentration, pH, and solvent properties.
Structural studies have shown that {\it in vitro}, A$\beta$ fibril
formation is preceded by formation of intermediates, spherical
oligomeric states and protofibrils (Walsh et al., 1997; Hartley et
al., 1999; Walsh et al., 1999; Kirkitadze et al., 2001; Yong et
al., 2002).  Structural studies on oligomeric states are in a less
advanced stage compared to those in fibrils.  The nature and structure
of different oligomeric states may depend crucially on the specific
amino acid sequence of the peptide (Nilsberth et al., 2001).  The
A$\beta$ plaques in AD brain are predominantly comprised of two A$\beta$
alloforms, A$\beta$(1-40) and A$\beta$(1-42).  Despite the relatively
small structural difference between these two alloforms, they display
distinct behavior, with A$\beta$(1--42) being a predominant component of
parenchymal plaques (Suzuki et al., 1994; Iwatsubo et al., 1994;
Gravina et al., 1995), associated with both early onset AD (Scheuner et
al., 1996; Golde et al., 2000) and increased risk for AD (Weggen et
al., 2001).  The cause of the clinical differences between the two
alloforms is still unknown.  Recent experiments have shown that {\it in
vitro} A$\beta$(1--40) and A$\beta$(1--42) oligomerize through distinct
pathways with A$\beta$(1--42) forming spherical paranuclei which further
assemble into higher order oligomers (Bitan et al., 2001; Bitan et al.,
2003a; Bitan et al., 2003b).

Several studies found stable soluble A$\beta$ low molecular weight
oligomers (Barrow and Zagorski, 1991; Barrow et al., 1992; Zagorski
and Barrow, 1992; Soreghan et al., 1994; Shen and Murphy, 1995;
Podlisny et al., 1995; Roher et al., 1996; Kuo et al., 1996;
Garzon-Rodriguez et al., 1997; Xia et al., 1995; Enya et al.,
1999; Funato et al., 1999; Huang et al., 2000).  Low molecular
weight oligomers were found in culture media of Chinese hamster ovary
cells expressing endogenous or mutated genes (Podlisny et al., 1995;
Xia et al., 1995).  A$\beta$(1--40) and A$\beta$(1--42) oligomers,
specifically dimers, were isolated from human control and AD brains (Kuo
et al., 1996; Kuo et al., 1998; Enya et al., 1999; Funato et al.,
1999).  Dimers and trimers of A$\beta$ were isolated from neuritic and
vascular amyloid deposits and dimers were shown to be toxic to neurons
in the presence of microglia (Roher et al., 1996).  Experiments on
synthetic A$\beta$ peptides (Garzon-Rodriguez et al., 1997; Podlisny et
al., 1998) showed that soluble A$\beta$(1--40) exists as a stable dimer
at physiological concentrations which are well below the critical
micelle concentration (Soreghan et al., 1994).

It has been shown that the $\beta$-sheet content of A$\beta$ depends
strongly upon the solvent in which the peptide is dissolved (Shen and
Murphy, 1995).  Various experimental studies (Barrow and Zagorski,
1991; Barrow et al., 1992; Zagorski and Barrow, 1992; Shen and
Murphy, 1995) indicate that soluble A$\beta$ has substantial
$\beta$-sheet content.  Huang et al. (Huang et al., 2000) reported on
two types of soluble oligomers of A$\beta$(1-40) which were trapped and
stabilized for an extended period of time: the first type was a mixture
of dimers and tetramers with irregular secondary structure and the
second type corresponded to larger spherical particles with
$\beta$-strand structure.  Despite some discrepancies in the
experimental results, the studies mentioned above suggest that
dimerization may be the initial event in amyloid aggregation and thus
dimers may be fundamental building blocks for further fibril assembly.

Experimental methods, such as circular dichroism, nuclear magnetic
resonance and electron microscopy, provide only limited information on
the structure of intermediate oligomeric states.  Therefore, there is a
motivation to develop new computational approaches to determine the
exact conformation of oligomers at the atomic level and track the exact
pathway from individual monomer peptides to oligomers and protofibrils
in fast and efficient ways.  With the dramatic increase of computer
power in recent decades, it has become possible to study the behavior of
large biological molecular systems by Monte Carlo and molecular dynamics
(MD) simulations (Dinner et al., 2002; Fersht and Daggett, 2002;
Karplus and McCammon, 2002; Thirumalai et al., 2002; Plotkin and
Onuchic, 2002; Mendes et al., 2002; Mirny and Shakhnovich, 2001;
Bonneau and Baker, 2001; Dill, 1999; Levitt et al., 1997; Wolynes
et al., 1996; Snow et al., 2002; Vorobjev and Hermans, 2001; Zhou
and Karplus, 1997).  However, traditional all-atom MD with realistic
force fields in a physiological solution currently remains
computationally unfeasible.  An aggregation process as allowed by
all-atom MD can only be studied on time scales of up to $10^{-7}$s using
such advanced technologies as worldwide distributed computing (Snow et
al., 2002; Zagrovic et al., 2002).  However, {\it in vivo} and {\it in
vitro} studies suggest that the initial stages of oligomerization occur
on a time scale of one second (Bitan et al., 2003a), while further
aggregation into protofibrillar and fibrillar aggregates may span hours
(Kayed et al., 2003).

Here we conduct a two--step study of A$\beta$ dimer conformations and
their stability using a computationally efficient algorithm combined
with a coarse--grained peptide model for A$\beta$.  We apply a
four--bead model for A$\beta$ peptide to study monomer and dimer
conformations of A$\beta$(1--42) peptide (Ding et al., 2003).  We use
fast and efficient discrete molecular dynamics (DMD) simulations
(Dokholyan et al., 1998; Smith and Hall, 2001a).  The DMD method
allows us to find and study a large variety of dimer conformations
starting from initially separated monomers without secondary structures.
Our coarse--grained model combined with the DMD method predicts 10
different planar $\beta$-strand dimer conformations.  In the second
step, we estimate the free energy of A$\beta$(1--42) and A$\beta$(1--40)
dimeric conformations in a stability study using all-atom MD simulations
with explicit water and well-established force fields.  This second step
enables us to estimate the free energy of different dimeric
conformations and to compare the free energies of A$\beta$(1--42) and
A$\beta$(1--40) for each of the dimer peptides.  Our results suggest
that A$\beta$ oligomerization is not accompanied by the formation of
stable planar $\beta$-strand A$\beta$ dimers, and that such dimers of
both A$\beta$(1-42) and A$\beta$(1-40) are equally unlikely to represent
stable oligomeric forms.

%======================================================================
\section{Methods}

\subsection{Discrete molecular dynamics simulations} 

In a DMD simulation, pairs of particles interact by means of
spherically-symmetric potentials that consist of one or more
square wells.  Within each well the potential is constant. 
Consequently, each pair of particles moves with constant
velocity until they reach a distance at which the potential
changes.  At this moment a collision occurs and the two
particles change their velocities instantaneously while
conserving the total energy, momentum, and angular momentum. 
There are three main types of collisions.  The simplest is
when particles collide at their hard--core distance, the sum
of the particle radii.  In this case, the particles collide
elastically, and their kinetic energy before and after the
collision is conserved.  In the second case, the particles
enter a potential well of depth $\Delta U$.  In this case,
their total kinetic energy after the collision increases by
$\Delta U$, their velocities increase, and there is a change
in their trajectories.  In the third case, particles exit a
potential well of depth $\Delta U$.  Here, total kinetic
energy after the collision decreases by $\Delta U$.  If the
total kinetic energy of the particles is greater than $\Delta
U$, they escape the well.  If their total kinetic energy is
smaller than $\Delta U$, the particles can not escape and
simply recoil from the outer border of the well inwards.  At
low temperatures, which correspond to low average particle
kinetic energies, particles whose potentials are attractive
thus have a tendency to remain associated with each other.

DMD, unlike traditional continuous MD, is event-driven and as such it
requires keeping track of particle positions and velocities only at
collision times, which have to be sorted and updated.  It can be shown
that the speed of the most efficient DMD algorithm is proportional to $N
\ln N$, where $N$ is the total number of atoms (Rapaport, 1997).  In
addition, the speed of the algorithm decreases linearly with the number
of discontinuities in the potential and particle density.  In our DMD
simulations the solvent is not explicitly present, which reduces the
number of particles in the system.  Consequently, the DMD method is
several orders of magnitude faster than the traditional continuous MD.
The DMD simulation method has been so far successfully applied to
simulate protein folding (Zhou and Karplus, 1997; Dokholyan et al.,
1998; Ding et al., 2002a; Borreguero et al., 2002) and aggregation
(Smith and Hall, 2001a; Smith and Hall, 2001b; Ding et al., 2002b;
Ding et al., 2003).  In simulating protein folding and aggregation,
coarse--grained models of proteins have been introduced.  In a
coarse--grain model the number of atoms per amino acid is reduced to
one, two or four, which further speeds up the DMD simulation.  While
traditional continuous all-atom MD can simulate events on time scales of
nanoseconds, the DMD method combined with a coarse--grained protein
model can easily reach time scales of seconds or more, which is long
enough to study oligomer formation of up to 100 A$\beta$ peptides.

\subsection{A coarse--grained model for A$\beta$ peptide}

The A$\beta$ peptide is derived from its larger amyloid precursor
protein by sequential proteolytic cleavages.  In amyloid plaques, the
two most common forms of the A$\beta$ peptide are A$\beta$(1--40) and
A$\beta$(1--42).  The amino acid sequence of A$\beta$(1--42) is
$$DAEFRHDSGYEVHHQKLVFFAEDVGSNKGAIIGLMVGGVVIA.$$ The amino acid sequence
of A$\beta$(1--40) is the same as that of A$\beta$(1--42), but shorter
by two amino acids at the C-terminus, Ile and Ala ($IA$ in the above
sequence).

In our DMD simulations we apply the simplest version of the four--bead
model (Takada et al., 1999; Smith and Hall, 2001a; Smith and Hall,
2001b; Ding et al., 2003) for A$\beta$ peptide.  In this model, each
amino acid in the peptide is replaced by at most four ``beads''.  These
beads correspond to the atoms comprising the amide nitrogen $N$, the
alpha carbon $C_{\alpha}$ and prime-carbon $C'$.  The fourth bead,
representing the amino acid side-chain atoms, is placed at the center of
the nominal $C_{\beta}$ atom.  Due to their lack of side-chains, the six
glycines in A$\beta$ (positions 9, 25, 29, 33, 37 and 38) are
represented by only three beads.  Two beads that form a permanent bond
can assume any distance between the minimum and maximum bond length.  In
addition to permanent bonds between the beads, the model introduces
constraints between pairs of beads which do not form permanent bonds.
These constraints are implemented in order to account for the correct
peptide backbone geometry.  The hard--core radii, minimum and maximum
bond lengths, constraints' lengths and their corresponding standard
deviations are either calculated from distributions of experimental
distances between pairs of these groups found in about 7700 folded
proteins with known crystal structures (Protein Data Bank), or chosen
following the standard knowledge of the geometry of the peptide backbone
(Creighton, 1993).  The values of all these parameters have been
reported previously (Ding et al., 2003).

To account for the hydrogen bonding that normally occurs in proteins
between the carbonyl oxygen of one amino acid and the amide hydrogen of
another amino acid, the coarse--grained model implements a bond between
the nitrogen of the $i-th$ amino acid, $N_i$, and the carbon of the
$j-th$ amino acid, $C'_j$, as introduced previously (Ding et al., 2003).
The planar geometry of the hydrogen bond is modeled by introducing
auxiliary bonds between the left and the right neighboring beads of
$N_i$ and $C'_j$.  The hydrogen bond between $N_i$ and $C'_j$ will form
only if all six beads are at energetically favorable distances.  Once
the hydrogen bond is formed, it can break due to thermal fluctuations,
which can cause energetically unfavorable distances among the six beads
involved in the hydrogen bond.  When amino acids $i$ and $j$ belong to
the same peptide, they can form a hydrogen bond only if at least three
amino acids exist between them (to satisfy the 180$^\circ$ NH-CO bond
angle).  A more detailed description of the hydrogen bond implementation
has been given elsewhere (Ding et al., 2003).  Our current
implementation differs slightly from before (Ding et al., 2003): one of
the auxiliary bonds, namely the auxiliary bond between $N_i$ and the
bead $N$ (the nearest neighbor of $C'_j$) has a shorter equilibrium
distance: instead of $5.10 \pm 0.31 \AA$ used previously (Ding et al.,
2003), it is $4.70 \pm 0.08 \AA$.  This slight change in the auxiliary
bond length stabilizes the $\beta$-hairpin monomer conformation in our
model as described in the results section.

\subsection{All-atom molecular dynamics in an explicit solvent}

Next we detail how we use all-atom MD simulations in an explicit solvent
to compute the conformational free energy of A$\beta$(1--40) and
A$\beta$(1--42) monomer and dimer conformations.

\subsubsection{Preparation of peptide conformations}

For A$\beta$(1--40) monomer peptide structures we use ten NMR structures
with coordinates (Coles et al., 1998) (ID code name 1BA4 of Protein
Database Bank).  For each of these ten A$\beta$(1--40) monomer
structures, we construct a corresponding A$\beta$(1--42) monomer
structure by adding two residues, Ile and Ala, to the C-terminus of the
peptide using the SYBYL (Tripos Inc.) molecular modeling package.

All A$\beta$(1-42) dimer conformations in this study are generated by
DMD simulations using the four--bead model as described above.  Dimer
conformations, initially in the four--bead representation, are converted
into all--atom representation by using all--atom template amino acids.
These templates are superposed onto the coarse--grained amino acids such
that the four beads of the coarse--grained model coincide with the $N$,
$C_\alpha$, $C'$ and $C_\beta$ groups of the all--atom template amino
acids.  The new template coordinates with increased number of degrees of
freedom are optimized for preserving backbone distances as well as
formation of peptide planes.  This optimization is performed by rotating
the template amino acid along two axes, the $C_\alpha$--$N$ and the
$C'$--$C_\alpha$ axes using a Monte Carlo algorithm.  After positioning
the backbone atoms, the positions of side--chain atoms are determined by
avoiding steric collisions with the backbone and other neighboring
residues.  Positioning of side--chain atoms also follows a Monte Carlo
algorithm, during which side--chain atoms are rotated sequentially along
the $C_\alpha$--$C_\beta$ axis and the $C_\beta$--$C_\gamma$ axis in order
to find the optimal combination of axes angles that prevent collisions.
The backbone structure of the resulting peptide remains very close to
the initial structure of the peptide: the lengths of bonds and
constraints after the conversion are within the limits given by our
coarse--grained model.  The A$\beta$(1-40) dimer conformations
corresponding to each A$\beta$(1-42) dimer are constructed by disposing
of the last two amino acids of each A$\beta$(1-42) dimer conformation.

\subsubsection{Calculation of the conformational free energy in water}

We estimate conformational free energies of monomers and dimers in a
water environment using all--atom MD simulations.  All MD calculations
are performed using the Sigma MD program (Hermans et al., 1994) with
CEDAR force fields (Ferro et al., 1980; Hermans et al., 1984).  We
complete the all--atom reconstruction from above by adding hydrogen
atoms and solvating the peptide(s) in a SPC water model bath (Berendsen
et al., 1981).  We use periodic boundary conditions on a cubic box whose
sides extended 12$\AA$ beyond the leading edge of the peptide(s) on all
sides.  The MD method consists of two stages, equilibration and
production (Vorobjev et al., 1998; Vorobjev and Hermans, 1999;
Vorobjev and Hermans, 2001; Leach, 2001).  Equilibration allows both
the peptides and water to relax to a local energy minimum.  The steps of
equilibration, (1)-(7), and the production (8) are as follows:

\begin{description}
\item{(1)} {minimize the energy of the water---peptides are kept
immobile;}

\item{(2)} {perform MD simulations on the water using the NVT
ensemble at a temperature $T=200K$ for 96$ps$ (the time step
is 1$fs$)---peptides are kept immobile;}

\item{(3)} {minimize the energy of the water a second
time---peptides are kept immobile;}

\item{(4)} {minimize the energy of the peptides---water
molecules are kept immobile;}

\item{(5)} {perform MD simulations of the peptide using the
NVT ensemble at a temperature ($T=100K$)---water molecules
are kept immobile;}

\item{(6)} {minimize the energy of the peptides a second
time---water molecules are kept immobile;}

\item{(7)} {minimize the energy of the peptides and water
molecules simultaneously;}

\item{(8)} {perform the production run, i.e., unconstrained MD
simulations on the peptides and water using the NPT ensemble
at $T=300K$ and $P=1atm$ for 196$ps$.}

\end{description}

At steps (1), (3), (4), (6) and (7) we use the steepest descent energy
minimization method.  During steps (2) and (5), which are parts of
equilibration, peptide(s) and water coordinates have to reach a local
energy minimum for the given force field and with respect to each other.
The temperatures are kept low so that there are no conformational
changes.

During the production run (8) we maintain constant temperature and
pressure by Berendsen coupling (Berendsen et al., 1984) and calculate
electrostatic forces using the particle--mesh Ewald procedure (Darden et
al., 1993).  We record a snapshot of the configuration every picosecond.
We calculate the free energy for each conformation by averaging the
instantaneous free energy for each of the 196 snapshots.  Each of these
snapshots represent a microconfiguration.  We calculate the free energy
for each configuration by the ES/IS method (Vorobjev and Hermans, 1999),
which uses an explicit solvent simulation with an implicit solvent
continuum model:
\begin{equation}
G_{A} = \langle U_{m}(x) \rangle_{A} + \langle W(x) \rangle_{A} - TS_{conf,A},
\label{free_energy_definition}
\end{equation}
where $\langle ... \rangle_{A}$ denotes an average over all recorded
microconfigurations of the conformation A, $U_m$ is the intra--protein
conformational energy, and $S_{conf,A}$ is the entropy of conformation
A.  The intra--protein conformational energy, $U_m$, is a sum of two
terms: one is the short--range energy of packing, $U_{m,pack}$, and the
other is the electrostatic energy due to coulombic interactions,
$U_{m,coul}$.  The solvation free energy, $W(x)$, is the sum of three
terms: the first one, $G_{cav}$, is the energy required to form a cavity
in the solvent; the second one, $G_{s,vdw}$, is a contribution of the
van der Waals interactions between solvent and protein; and the third
one, $G_{pol}$, is a contribution of the electrostatic polarization of
the solvent and polar components of the solute.  Thus the above equation
becomes
\begin{equation}
G_{A}  = \langle U_{m,pack} \rangle_{A} + \langle U_{m,coul} \rangle_{A} 
       + \langle G_{cav} \rangle_{A} 
       + \langle G_{s,vdw} \rangle_{A} 
       + \langle G_{pol} \rangle_A - T S_{conf,A} \, .
\label{free_energy_calculation}
\end{equation}
We determine $U_m$ and $G_{s,vdw}$ from the MD trajectory, calculate
$G_{cav}$ as proportional to the accessible surface area for a given
micro-configuration, and evaluate $G_{pol}$ using an implicit model for
the solvent (in our case water) as described elsewhere (Vorobjev and
Hermans, 1999).

%======================================================================
\section{Results}

\subsection{Characterization of monomer conformations}

The secondary structure of both A$\beta$(1--40) and A$\beta$(1--42)
monomer peptides, as determined by NMR conformational studies in an
apolar environment that mimics the lipid phase of membranes, is
predominantly $\alpha$-helical.  Two $\alpha$-helical regions exist at
residues 8-25 and 28-38, and these regions are separated by a flexible
hinge.  The rest of the peptide adopts random coil-like conformation
(Coles et al., 1998; Crescenzi et al., 2002).

In order to characterize the monomer conformations in our
coarse--grained model, we calculate an average potential energy in
dependence on the temperature.  The energy unit corresponds to the
potential energy of one hydrogen bond in our model, so that the absolute
average of the potential energy is equal to the average number of
hydrogen bonds in the monomer conformation, and the temperature unit is
equal to the energy unit.  At each temperature $T$, $0.080 < T < 0.155$,
we perform $35 \times 10^6$ time steps long simulation runs.  We start
each run with an initial conformation equal to the observed NMR
conformation with predominantly $\alpha$-helical secondary structure
(Crescenzi et al., 2002).  The first $15 \times 10^6$ steps we allow for
equilibration, whereas we calculate the time average of the potential
energy $\langle E \rangle$ over the last $20 \times 10^6$ time steps.

Our monomer peptide experiences a structural transition from a
predominantly $\alpha$-helix conformation into a $\beta$-strand
conformation at $T_{\alpha,\beta} = 0.107 \pm 0.002$, in agreement with
previous work (Ding et al., 2003).  At a higher temperature,
$T_{\beta,RC} = 0.128 \pm 0.002$, the monomer undergoes a transition
from a $\beta$-strand into a random coil conformation with no particular
secondary structure.  Between $T_{\alpha,\beta}$ and $T_{\beta,RC}$ our
simulations show various types of $\beta$-strand rich conformations.

% While in the original coarse--grained model (Ding et al., 2003)
% the temperature range over which we observe $\beta$-strand
% monomer conformation is $0.130 < T < 0.150$, in the present
% model the corresponding temperature range is shifted to a
% lower temperatures, $0.107 < T < 0.128$.  This temperature
% shift is easy to understand if one takes into account that
% one of the hydrogen-bond constraints (between 2 N atoms
% within a hydrogen bond) has less flexibility because of a
% smaller width of the square well potential, $0.16\AA$, as
% compared to the original $0.62\AA$.  The more important
% difference between the present and the
% original (Ding et al., 2003) model is that in the present model
% there is a relatively wide temperature range over which we
% find a $\beta$-hairpin monomer conformation with a well
% defined $\beta$-turn as described below, a feature which
% can not be recovered within the original model.

At temperatures $T$, $T < 0.107$, we observe an $\alpha$-helix
conformation which is consistent with the observed solution monomer
conformation in an apolar microenvironment (Crescenzi et al., 2002).
This conformation (Fig.~1a) has a random
coil-like tail about 10 amino acids long at the N-terminus and another
random coil-like tail about 2-4 amino acids long at the C-terminus.  At
residues 11-40 there are two $\alpha$ helices, separated by a hinge at
residues 25-28.  The average potential energy of this conformation is
$\langle E \rangle = -28 \pm 2$.  At temperatures $0.107 < T < 0.117$,
we observe various $\beta$-strand conformations, mostly with two or
three $\beta$-turns, corresponding to 3 or 4 $\beta$-strands
(Figs.~1b-c).  The average potential energy
of these conformations is $\langle E \rangle = -17 \pm 1$.  The
$\beta$-hairpin conformation, i.e. a 2-$\beta$-strand conformation with
one $\beta$-turn, shown in Fig.~1d, is found
as a predominant conformation at temperatures $T$, $0.117 < T < 0.126$.
This conformation is characterized by a random coil tail at residues 1-9
and by a well defined and localized $\beta$-turn which is positioned at
residues 23-28.  The average potential energy of this conformation is
$\langle E \rangle = -13 \pm 1$.

The observed $\beta$-turn between residues Asp 23 and Lys 28 is in
agreement with recent NMR studies of A$\beta$ fibrillar structure
(Petkova et al., 2002).  In the following, we provide an empirical
explanation for the occurrence of this well--defined $\beta$-turn in our
model.  We hypothesize that within our model the occurrence of a
$\beta$-turn at resides 23-28 is induced by the particular location of
the six glycines in the A$\beta$(1-42) peptide.  In order to test this
hypothesis, we replace all glycines within the A$\beta$(1-42) peptide
with alanines and perform simulation runs as described above.  Our
results (Fig.~2) show the probability for the amino
acid at a certain position to be part of a $\beta$-turn both for the
original A$\beta$ peptide model (with glycines) and the one without
glycines (42 amino acids long polyalanine chain).  The results show that
(i) the presence of glycines on average shifts the center of the
$\beta$-turn from residue 20-22 for the chain with no glycines to 25-27
for the chain with six glycines, and (ii) the probability distribution
in the presence of glycines is strongly peaked at residues 25-27, which
makes these three residues part of the $\beta$-turn with more than 95\%
probability and thus $\beta$-turn is well--defined.

The residues 25-27 of the A$\beta$ peptide correspond to glycine, serine
and asparagine, the residues which have according to the classical
phenomenological approach of Chou and Fasman (Chou and Fasman, 1974) the
highest probability to be within a $\beta$-turn.  In our coarse--grained
model the occurrence of the $\beta$-turn at 23-28 can be understood as a
consequence of two tendencies: (1) a tendency to maximize the number of
hydrogen bonds, which prefers a $\beta$-turn at the middle of the
peptide chain, centered at resides 20-22; and (2) a tendency of 6
glycines to be associated with more flexibility, thus a $\beta$-turn.
Consequently, the center of the $\beta$-turn is shifted from residues
20-22 to residues 25-27, and is well-defined.

\subsection{Planar $\beta$-strand dimer conformations of A$\beta$(1--42)}

We investigate next dimer formation of A$\beta$(1--42) peptides.  The
initial monomer conformations are taken from the Protein Database Bank
and correspond to the observed NMR structures of A$\beta$(1-42) monomers
in an apolar environment (Crescenzi et al., 2002).  In order to obtain
different starting random coil conformations, we place two monomers with
mostly $\alpha$-helical secondary structure in a cubic box with a side
length of $100 \AA$.  The centers of masses of the two monomers are
initially about $50 \AA$ apart and their orientations parallel.  Next,
we heat the system up to a temperature $T=0.50$, which is far above the
observed $T_{\beta,RC}$ temperature.  The $\alpha$-helical secondary
structure of individual monomer peptides is dissolved in about $200$
simulations steps, producing two peptides with different random coil
conformations.  We use many similarly generated pairs of peptides with
random coil conformations as initial configurations in our study of
dimer formation.  Dimer formation runs are done at a constant
temperature and volume.  We perform 20 runs at a fixed temperature.
Each run is $20 \times 10^6$ time steps long.  In this way we explore
temperatures $T=0.120$, $0.125$ and $0.130$.

From the above simulations we find six possible dimer conformations with
the following characteristics: (i) each peptide in a dimer is in a
$\beta$-hairpin conformation with two $\beta$-strands and (ii) all four
$\beta$-strands (two per peptide) are planar.  We name those dimers
according to the inner two strands of the dimer (each strand is either
closer to the N-terminus or the C-terminus and the two inner strands are
either parallel or antiparallel): NN-parallel, NC-parallel, CC-parallel,
NN-antiparallel, NC-antiparallel, CC-antiparallel.  These conformations
are schematically presented in Fig.~3a-f.  We find
four additional dimer conformations with the characteristic (ii)
described above.  Only the inner peptide has also the characteristic
(i), while the outer peptide is bent around the inner one, forming a
``nest''.  We term them nested parallel, nested antiparallel,
anti-nested parallel and anti-nested antiparallel (in anti-nested
conformations the termini of the two peptides are in the opposite
directions).  They are shown in Fig.~3g-j.

At $T=0.12$, we find NC-parallel and NC-antiparallel conformations each
in 3 out of 20 runs.  The conformations NN-parallel, CC-parallel,
CC-antiparallel, nested antiparallel, anti-nested parallel and
anti-nested antiparallel, each occur in 2 out of 20 runs.  The
conformations NN- antiparallel and nested parallel each are found in 1
out of 20 runs.  At $T=0.13$, the most common dimer peptide conformation
is NC-parallel (occurring in 8 out of 20 runs) and the next most common
conformation is NN-parallel (occurring in 5 out of 20 runs).  We find
the NC-antiparallel conformation in 3 out of 20 runs.  There are four
more conformations found, each in 1 out of 20 runs: NN-antiparallel,
CC-parallel, CC-antiparallel and a nested-antiparallel conformation.

Our dimer simulation runs at temperatures $T \ge 0.14$ show no
dimerization within the first $20 \times 10^6$ simulation steps, even
though typically one of the two peptides adopts one of the
$\beta$-strand conformations.  We thus conclude that at temperatures
$T>0.14$ there is no dimerization.  At temperature $T=0.11$, we observe
a large number of different planar and non--planar $\beta$-strand dimer
conformations, which are a mixture of 2-, 3-, 4-$\beta$-strand
conformations.  At temperatures $0.08 < T < 0.11$, the dimer
conformations are an amorphous mixture of $\beta$-strand and
$\alpha$-helical secondary structure.  All these are omitted from the
present all-atom free energy calculation study.

%======================================================================
\subsection{Free energy calculations:  A$\beta$(1-40) versus
A$\beta$(1-42) monomer conformations.}

In order to validate the free energy calculation method, we first
analyze monomer peptides of A$\beta$(1-40) and A$\beta$(1-42).  We
choose 10 different NMR A$\beta$(1-40) monomer structures (Coles et al.,
1998).  The secondary monomer structure is mostly $\alpha$-helical,
similar to Fig.~1a.  To each of these
structures we add 2 amino acids, Ile and Ala, to find the corresponding
A$\beta$(1-42) monomer conformation.  The estimate free energies are
presented in Table I and show that all the monomer conformations, of
A$\beta$(1-40) and A$\beta$(1-42), have on average the same
conformational free energy, $-1034.68 \pm 17.75 kcal/mol$ and $-1029.47
\pm 10.80 kcal/mol$, respectively.  These results show that addition of
two amino acids to the C-terminus does not alter the conformational free
energy in a water environment at physiological conditions.

\subsection{Stability analysis of A$\beta$(1-40) and
(1-42) dimer conformations.}

The planar $\beta$-strand dimer conformations predicted by our
coarse--grained model (Fig.~3) are tested for
stability in our all--atom MD simulations in an explicit water
environment at atmospheric pressure and room temperature.  From ten
different A$\beta$(1-42) dimer conformations, we create the
corresponding A$\beta$(1-40) dimers by deleting the last two amino acids
at the C-terminus.  For each stable dimer configuration we then
calculate the free energy as described in the Methods section.  The free
energies of all the stable dimers are presented in Table II.  One dimer
conformation, e.g. nested antiparallel of A$\beta$(1-40), is determined
to be only marginally stable and does not allow for the free energy
calculation.  Columns 3 and 6 of Table II represent the free energy
differences, $\Delta$G$_{A\beta-40}$ and $\Delta$G$_{A\beta-42}$,
between a dimer conformation and two monomer conformations.  For all
stable dimer conformations $\Delta$G is positive, indicating that in a
water environment, planar $\beta$-strand dimer conformations are
energetically unfavorable compared to $\alpha$-helical monomer peptide
conformations.  The average conformational free energies of
A$\beta$(1-40) and A$\beta$(1-42) dimers are $-2000.81 \pm 46.94
kcal/mol$ and $-1967.63 \pm 52.85 kcal/mol$, respectively.  Although
A$\beta$(1-40) dimers have on average lower conformational free energies
than A$\beta$(1-42) dimers, the difference is not statistically
significant.

%======================================================================
\section{Discussion and Conclusions}

We introduce a coarse--grained peptide model for the A$\beta$ peptide in
order to study A$\beta$ dimer formation.  Our model predicts a
thermally--induced conformational change between a predominantly
$\alpha$-helix to a predominantly $\beta$-strand monomer peptide.  The
prediction of our model is indirectly supported by recent experiments
(Gursky and Aleshkov, 2000) on temperature--dependence of the A$\beta$
conformation in aqueous solutions, which show that thermally induced
coil to $\beta$-strand transition is not coupled to aggregation and can
occur at the level of monomers or dimers.  In a temperature range below
the structural transition into a random coil our model A$\beta$ monomer
peptide adopts a $\beta$-hairpin conformation with a $\beta$-turn
between Asp 23 and Lys 28.  The presence of this $\beta$-turn is
consistent with a structural model for A$\beta$ fibrils based on solid
state NMR experimental constraints (Petkova et al., 2002).

Within our coarse--grained model we study A$\beta$ dimer formation,
which may be a pathway to higher oligomer and protofibril formation.  In
our model A$\beta$ dimers are formed as a consequence of hydrogen-bond
interactions between residues.  Our model predicts that dimer
conformations are $\beta$-sheet-like planar structures.  We show by
using all-atom simulations that these planar, $\beta$-sheet-like dimer
conformations are energetically unfavorable compared to the
$\alpha$-helical monomer conformations in water environment.  Moreover,
the free energy comparison of A$\beta$(1-40) and A$\beta$(1-42) dimer
conformations shows that there is no significant free energy difference
between these two alloforms, thus suggesting that A$\beta$
oligomerization is not accompanied by the formation of stable planar
$\beta$-strand A$\beta$ dimers.
  
Planar $\beta$-strand A$\beta$ dimers as predicted by our
coarse--grained model cannot account for experimentally observed
differences in A$\beta$ oligomer formation between A$\beta$(1-40) and
A$\beta$(1-42) alloforms (Bitan et al., 2003a).  It is not understood
yet at which stage of oligomer formation those differences occur and
what is the exact mechanism which drives the two alloforms along
different pathways.  In order to account for oligomer formation
differences between the two A$\beta$ alloforms, our coarse--grained
model may need to include other interactions between the residues, in
particular the ones that originate in polar versus apolar character of
the side--chains as suggested by recent experiments (Bitan et al.,
2003c; Bitan et al., 2003b).

%======================================================================
\acknowledgments 

This work was supported by the Memory Ride Foundation.  NVD
acknowledges the support of UNC Research Council grant and
MDA grant (MDA3702).  We thank J.  Hermans, E.  I. 
Shakhnovich and D.  B.  Teplow for valuable advice,
discussions and critical reviews of the manuscript. 

%======================================================================
\newpage
%\begin{thebibliography}{99}

\subsection*{References}
{\parindent=0pt
\parskip=10pt

Antzutkin, O.N., R.D. Leapman, J.J. Balbach, and R. Tycko. 2002.
Supramolecular structural constraints on Alzheimer's beta-amyloid
fibrils from electron microscopy and solid-state nuclear magnetic
resonance.  Biochemistry 41:15436--15450.

Antzutkin, O.N., J.J. Balbach, and R. Tycko. 2003.  Site-specific
identification of non-beta-strand conformations in Alzheimer's
beta-amyloid fibrils by solid-state NMR.  Biophys. J. 84:3326--3335.

Balbach, J.J., A.T. Petkova, N.A. Oyler, O.N. Antzutkin, D.J. Gordon,
S.C. Meredith, and R.  Tycko. 2002.  Supramolecular structure in
full-length Alzheimer's $\beta$-amyloid fibrils: evidence for a parallel
$\beta$-sheet organization from solid-state nuclear magnetic
resonance. Biophys. J. 83:1205--1216.

Barrow, C.J. and M.G. Zagorski. 1991.  Solution structures of beta
peptide and its constituent fragments: relation to amyloid deposition.
Science 253:179--182.

Barrow, C.J., A. Yasuda, P.T.M. Kenny, and M.G.  Zagorski. 1992.
Solution conformations and aggregational properties of synthetic amyloid
beta-peptides of Alzheimer's disease. Analysis of circular dichroism
spectra.  J. Mol. Biol. 225:1075--1093.

Berendsen, H.J.C., J.P.M. Postma, W.F. van Gunsteren, and
J. Hermans. 1981.  Interaction models for water in relation to protein
hydration. In Intermolecular Forces. B.  Pullman, editor. Reidel,
Dordrecht. 331--342.

Berendsen, H.J.C., J.P.M. Postma, W.F. van Gunsteren, A, Dinola, and
J.R. Haak. 1984.  Molecular-dynamics with coupling to an external bath.
J. Chem. Phys. 81:3684--3690.

Bitan, G., A. Lomakin, and D.B. Teplow. 2001.  Amyloid-$\beta$-protein
oligomerization: prenucleation interactions revealed by photo-induced
cross-linking of unmodified proteins.  J. Biol. Chem. 276:35176--35184.

Bitan, G., M.D. Kirkitadze, A. Lomakin, S.S. Vollers, G.B. Benedek, and
D.B. Teplow.  2003a.  Amyloid beta-protein (Ab) assembly: A beta 40 and
A beta 42 oligomerize through distinct pathways.
Proc. Natl. Acad. Sci. 100:330--335.

Bitan, G., B. Tarus, S.S. Vollers, H.A.  Lashuel, M.M. Condron,
J.E. Straub, and D.B. Teplow.  2003b.  A molecular switch in amyloid
assembly: Met$^{35}$ and amyloid beta-protein oligomerization.
J. Am. Chem. Soc. 125:15359--15365.

Bitan, G., S.S. Vollers, and D.B. Teplow.  2003c.  Elucidation of
primary structure elements controlling early amyloid beta-protein
oligomerization.  J. Biol. Chem. 278:34882-34889.

Bonneau, R. and D. Baker. 2001.  Ab initio protein structure prediction:
progress and prospects.  Annu. Rev. Bioph. Biom. 30:173--189.

Borreguero, J.M., N.V. Dokholyan, S.V.  Buldyrev, H.E. Stanley, and
E.I. Shakhnovich. 2002.  Thermodynamics and folding kinetic analysis of
the SH3 domain from discrete molecular dynamics.
J. Mol. Biol. 318:863--876.

Bucciantini, M., E. Giannoni, F.  Chiti, F. Baroni, L. Formigli,
J. Zurdo, N. Taddel, G.  Ramponi, C.M. Dobson, and M. Stefani. 2002.
Inherent toxicity of aggregates implies a common mechanism for protein
misfolding diseases. Nature 416:507--511.

Chou, P.Y. and G.D. Fasman. 1974.  Prediction of protein conformation.
Biochemistry 13:222--245.

Coles, M., W. Bicknell, A.A Watson, D.P. Fairlie, and D.J. Craik. 1998.
Solution structure of amyloid-beta peptide (1-40) in a water-micelle
environment: Is the membrane spanning domain where we think it is?
Biochemistry 37:11064-11077.

Creighton T.E. 1993. Proteins: Structures and Molecular Properties, 2nd
Edition.  Freeman and Company, New York.

Crescenzi, O., S. Tomaselli, R. Guerrini, S.  Salvatori, A.M. D'Ursi,
P.A. Temussi, and D. Picone. 2002.  Solution structure of the Alzheimer
amyloid $\beta$-peptide (1-42) in an apolar microenvironment: similarity
with a virus fusion domain.  Eur. J. Biochem. 269:5642--5648.

Dahlgren, K.N., A.M. Manelli, W. Blaine Stine Jr., L.K. Baker,
G.A. Krafft, and M.J. LaDu. 2002. Oligomeric and fibrillar species of
amyloid-$\beta$ Peptides differentially affect neuronal
viability. J. Biol. Chem. 277:32046--32053.

Darden, T., D.M. York, and L.G. Pedersen. 1993.  Particle mesh Ewald: an
N log(N) method for Ewald sums in large systems.
J. Chem. Phys. 98:10089--10092.

Dill, K.A. 1999.  Polymer principles and protein folding.
Prot. Sci. 8:1166--1180 1999.

Ding, F., N.V. Dokholyan, S.V.  Buldyrev, H.E. Stanley, and
E.I. Shakhnovich. 2002a.  Direct molecular dynamics observation of
protein folding transition state ensemble, Biophys. J. 83:3525--3532.

Ding, F., N.V. Dokholyan, S.V.  Buldyrev, H.E. Stanley, and
E.I. Shakhnovich. 2002b.  Molecular dynamics simulation of the SH3
domain aggregation suggests a generic amyloidogenesis mechanism.
J. Mol. Biol. 324:851--857.

Ding, F., J.M. Borreguero, S.V. Buldyrev, H.E.  Stanley, and
N.V. Dokholyan. 2003.  A mechanism for the $\alpha$-helix to
$\beta$-hairpin transition.  Proteins: Structure, Function, and Genetics
53:220--228.

Dinner, A.R., S.S. So, and M. Karplus. 2002.  Statistical analysis of
protein folding kinetics.  Adv. Chem. Phys. 120:1--34.

Dodart, J.-C., K.R. Bales, K.S. Gannon, S.J.  Greene, R.B. DeMattos,
C. Mathis, C.A. DeLong, S.  Wu, W. Wu, D.M. Holtzman, and
S.M. Paul. 2002.  Immunization reverses memory deficits without reducing
brain Abeta burden in Alzheimer's disease
model. Nat. Neurosci. 5:452--457.

Dokholyan, N.V., S.V. Buldyrev, H.E.  Stanley, and
E.I. Shakhnovich. 1998.  Molecular dynamics studies of folding of a
protein-like model.  Folding and Design 3:577--587.

El-Agnaf, O.M.A., D.S. Mahil, B.P. Patel, and B.M. Austen. 2000.
Oligomerization and toxicity of $\beta$-amyloid-42 implicated in
Alzheimer's disease. Biochem. Biophys. Res. Commun. 273:1003--1007.

El-Agnaf, O.M.A., S. Nagala, B.P. Patel, and B.M. Austen. 2001.
Non-fibrillar oligomeric species of the amyloid ABri peptide, implicated
in familial British dementia, are more potent at inducing apoptotic cell
death than protofibrils or mature fibrils. J. Mol. Biol. 310:157--168.

Enya, M., M. Morishima-Kawashima, M. Yoshimura, Y. Shinkai, K. Kusui,
K. Khan, D. Games, D. Schenk, S.  Sugihara, H. Yamaguchi, and
Y. Ihara. 1999.  Appearance of sodium dodecyl sulfate-stable amyloid
$\beta$- protein (A$\beta$) dimer in the cortex during aging.
Am. J. Pathol. 154:271--279.

Ferro, D.R., J.E. McQueen, J.T. McCown, and J. Hermans. 1980.  Energy
minimization of rubredoxin.  J. Mol. Biol. 136:1--18.

Fersht, A.R. and V. Daggett. 2002.  Protein folding and unfolding at
atomic resolution.  Cell 108:573--582 2002.

Funato, H., M. Enya, M. Yoshimura, M. Morishima-Kawashima, and
Y. Ihara. 1999.  Presence of sodium dodecyl sulfate-stable amyloid
$\beta$-protein dimers in the hippocampus CA1 not exhibiting
neurofibrillary tangle formation.  Am. J. Pathol. 155:23--28.

Garzon-Rodriguez, W., M. Sepulveda-Becerra, S. Milton, and
C.G. Glabe. 1997.  Soluble amyloid A$\beta$-(1--40) exists as a stable
dimer at low concentrations.  J. Biol. Chem. 272:21037--21044.

Golde, T.E., C.B. Eckman, and S.G. Younkin.  2000.  Biochemical
detection of A beta isoforms: implications for pathogenesis, diagnosis,
and treatment of Alzheimer's disease.  Biochem. Biophys. Acta Mol. Basis
Dis. 1502:172--187.

Gravina, S.A., L.B. Ho, C.B. Eckman, K.E.  Long, L. Otvos Jr.,
L.H. Younkin, N. Suzuki, and S.G.  Younkin.  1995.  Amyloid beta protein
(A beta) in Alzheimer's disease brain: biochemical and
immunocytochemical analysis with antibodies specific for forms ending at
A beta 40 or A beta 42(43).  J. Biol. Chem. 270:7013--7016.

Gursky, O. and S. Aleshkov. 2000.  Temperature-dependent $\beta$-sheet
formation in $\beta$-amyloid A$\beta_{1-40}$ peptide in water:
uncoupling $\beta$-structure folding from aggregation.
Biochim. Biophys. Acta 1476:93--102.

Hartley, D.M., D.M. Walsh, C.P. Ye, T.  Diehl, S. Vasquez,
P.M. Vassilev, D.B. Teplow, and D.J. Selkoe. 1999.  Protofibrillar
intermediates of amyloid $\beta$-protein induce acute
electrophysiological changes and progressive neurotoxicity in cortical
neurons.  J. Neurosci. 19:8876--8884.

Hermans, J., H.J.C. Berendsen, W.F. van Gunsteren, and
J.P.M. Postma. 1984.  A consistent empirical potential for water-protein
interactions.  Biopolymers 23:1513--1518.

Hermans, J., R.H. Yun, J. Leech, and D.  Cavanaugh. 1994. Sigma
documentation. University of North Carolina: \hfill \\
http://hekto.med.unc.edu:8080/HERMANS/software/SIGMA/index.html.

Hsia, A.Y., E. Masliah, L. McConlogue, G.-Q. Yu, G. Tatsuno, K. Hu,
D. Kholodenko, R.C. Malenka, R.A. Nicoll, and
L. Mucke. 1999. Plaque-independent disruption of neural circuits in
Alzheimer's disease mouse models. Proc. Natl. Acad.
Sci. 96:3228--3233.

Huang, T.H.J., D.-S. Yang, N.P. Plaskos, S.  Go, C.M. Yip, P.E. Fraser,
and A. Chakrabartty. 2000.  Structural studies of soluble oligomers of
Alzheimer $\beta$-amyloid peptide.  J. Mol. Biol. 297:73--87.

Humphrey, W., A. Dalke, and K. Schulten. 1996.  VMD: visual molecular
dynamics.  J. Molec. Graphics 14.1:33--38.

Iwatsubo, T., A. Odaka, N. Suzuki, H.  Mizusawa, N. Nukina, and
Y. Ihara. 1994.  Visualization of A beta 42(43) and A beta 40 in senile
plaques with end-specific A beta monoclonals: evidence that an initially
deposited species is A beta 42(43).  Neuron 13:45--53.

Karplus, M. and J.A. McCammon. 2002.  Molecular dynamics simulations of
biomolecules.  Nat. Struct. Biol. 9:646--652.

Kayed, R., E. Head, J.L. Thompson, T.M. McIntire, S.C.  Milton,
C.W. Cotman, and C.G. Glabe. 2003.  Common structure of soluble amyloid
oligomers implies common mechanisms of pathogenesis. Science
300:486--489.

Kirkitadze, M.D., M.M. Condron, and D.B. Teplow. 2001.  Identification
and characterization of key kinetic intermediates in amyloid b-protein
fibrillogenesis.  J. Mol. Biol. 312:1103--1119.

Kirkitadze, M.D., G. Bitan, and D.B. Teplow. 2002.  Paradigm shifts in
Alzheimer's disease and other neurodegenerative disorders: the emerging
role of oligomeric assemblies. J. Neurosci. Res. 69:567--577.

Klein, W.L., G.A. Krafft, and C.E. Finch. 2001.  Targeting small A beta
oligomers: the solution to an Alzheimer's disease conundrum? Trends in
Neurosciences 24:219--224.

Klein, W.L. 2002a.  A$\beta$ toxicity in Alzheimer's disease: globular
oligomers (ADDLs) as new vaccine and drug targets. Neurochemistry
International 41:345--352.

Klein, W.L. 2002b. ADDLs and protofibrils---the missing link?
Neurobiol. Aging 23:231--233.

Kuo, Y.-M., M.R. Emmerling, C. Vigo-Pelfrey, T.C. Kasunic,
J.B. Kirkpatrick, G.H. Murdoch, M.J.  Ball, and A.E. Roher. 1996.
Water-soluble A$\beta$ (N-40, N-42) oligomers in normal and Alzheimer
disease brains.  J. Biol. Chem. 271:4077--4081.

Kuo, Y.-M., S. Webster, M.R. Emmerling, N. De Lima, and
A.E. Roher. 1998.  Irreversible dimerization/tetramerization and
post-translational modifications inhibit proteolytic degradation of
A-$\beta$ peptides of Alzheimer's disease.  Biochimica et Biophysica
Acta 1406:291--298.

Lambert, M.P., A.K. Barlow, B.A Chromy, C. Edwards, R. Freed,
M. Liosatos, T. E. Morgan, I. Rozovsky, B. Trommer, K.L. Viola,
P. Wals, C. Zhang, C.E. Finch, G.A. Krafft, and W.L. Klein. 1998.
Diffusible, nonfibrillar ligands derived from Ab(1-42) are potent
central nervous system neurotoxins. Proc. Natl. Acad.
Sci. 95:6448--6453.

Leach, A.R. 2001. Molecular Modelling: Principles and Applications, 2nd
Edition. Prentice-Hall, New York.

Levitt, M., M. Gerstein, E. Huang, S. Subbiah, and J. Tsai. 1997.
Protein folding: the endgame.  Annu. Rev. Biochem. 66:549--579.

Malinchik, S.B., H. Inouye, K.E. Szumowski, and D.A. Kirschner. 1998.
Structural analysis of Alzheimer's $\beta$(1-40) amyloid: protofilament
assembly of tubular fibrils.  Biophys. J. 74:537--545.

Mendes, J., R. Guerois, and L. Serrano. 2002.  Energy estimation in
protein design.  Curr. Opin. Struc. Biol. 12:441--446.

Mirny, L. and E.I. Shakhnovich. 2001.  Protein folding theory: from
lattice to all-atom models.  Annu. Rev. Bioph. Biom. 30:361--396.

Mucke, L., E. Masliah, G.-Q. Yu, M. Mallory, E.M. Rockenstein,
G. Tatsuno, K. Hu, D. Kholodenko, K. Johnson-Wood, and
L. McConlogue. 2000. High-level neuronal expression of A$\beta$-(1-42)
in wild-type human amyloid protein precursor transgenic mice:
synaptotoxicity without plaque formation. J. Neurosci. 20:4050--4058.

Nilsberth, C., A. Westlind-Danielsson, C.B. Eckman, M.M. Condron,
K. Axelman, C. Forsell, C. Stenh, J. Luthman, D.B. Teplow, S.G. Younkin,
J.  N\"aslund, and L. Lannfelt. 2001.  The `Arctic' APP mutation (E693G)
causes Alzheimer's disease by enhanced A$\beta$ protofibril formation.
Nature Neuroscience 4:887--893.

Petkova, A.T., Y. Ishii, J.J. Balbach, O.N.  Antzutkin, R.D. Leapman,
F. Delaglio, and R. Tycko. 2002.  A structural model for Alzheimer's
$\beta$-amyloid fibrils based on experimental constraints from solid
state NMR.  Proc. Natl. Acad. Sci. 99:16742--16747.

Plotkin, S.S. and J.N. Onuchic. 2002.  Understanding protein folding
with energy landscape theory, Part I: basic concepts.
Rev. Biophys. 35:111--167.

Podlisny, M.B., B.L. Ostaszewski, S.L. Squazzo, E.H.  Koo, R.E. Rydell,
D.B. Teplow, and D.J. Selkoe. 1995.  Aggregation of secreted amyloid
$\beta$-protein into sodium dodecyl sulfate-stable oligomers in cell
culture.  J. Biol. Chem. 270:9564--9570.

Podlisny, M.B., D.M. Walsh, P. Amarante, B.L.  Ostaszewski,
E.R. Stimson, J.E. Maggio, D.B. Teplow, and D.J. Selkoe. 1998.
Oligomerization of endogenous and synthetic amyloid $\beta$-protein at
nanomolar levels in cell culture and stabilization of monomer by Congo
red.  Biochemisty 37:3602--3611.

Rapaport D.C. 1997. The Art of Molecular Dynamics Simulation. Cambridge
University Press, Cambridge.

Roher, A.E., M.O. Chaney, Y.-M. Kuo, S.D.  Webster, W.B. Stine,
L.J. Haverkamp, A.S. Woods, R.J.  Cotter, J.M. Tuohy, G.A. Krafft,
B.S. Bonnell, and M.R. Emmerling. 1996.  Morphology and toxicity of
A$\beta$ dimer derived from neuritic and vascular amyloid deposits of
Alzheimer's disease.  J. Biol. Chem. 271:20631--20635.

Scheuner, D., C. Eckman, M. Jensen, X.  Song, M. Citron, N. Suzuki,
T.D. Bird, J. Hardy, M.  Hutton, W. Kukull, E. Larson, E. LevyLahad,
M. Viitanen, E. Peskind, P. Poorkaj, G. Schellenberg, R. Tanzi,
W. Wasco, L. Lannfelt, D. Selkoe, and S. Younkin. 1996.  Secreted
amyloid beta-protein similar to that in the senile plaques of
Alzheimer's disease is increased in vivo by the presenilin 1 and 2 and
APP mutations linked to familial Alzheimer's disease.
Nat. Med. 2:864--870.

Selkoe, D.J. 1997. Alzheimer's disease: genotypes, phenotypes and
treatments. Science 275:630--631.

Serpell, L.C., C.C.F. Blake, and P.E. Fraser. 2000.  Molecular structure
of a fibrillar Alzheimer's A$\beta$ fragment.  Biochem. 39:13269--13275.

Shen, C.L.S. and M. Murphy.  1995.  Solvent effects on self-assembly of
beta-amyloid peptide.  Biophys. J. 69:640--651.

Smith, A. V. and C.K. Hall. 2001a.  $\alpha$-helix formation:
discontinuous molecular dynamics on an intermediate-resolution protein
model.  Proteins 4:344--360.

Smith, A.V. and C.K. Hall. 2001b.  Protein refolding versus aggregation:
computer simulations on an intermediate-resolution protein model.
J. Mol. Biol. 312:187--202.

Snow, C.D., N. Nguyen, V.S. Pande, and M. Gruebele. 2002.  Absolute
comparison of simulated and experimental protein-folding dynamics.
Nature 420:102--106.

Soreghan, B., J. Kosmoski, and C. Glabe. 1994.  Surfactant properties of
Alzheimer's A$\beta$ peptides and the mechanism of amyloid aggregation.
J. Biol. Chem. 269:28551--28554.

Suzuki, N., T.T. Cheung, X.-D. Cai, A.  Odaka, L. Otvos Jr., C. Eckman,
T.E. Golde, and S.G.  Younkin. 1994.  An increased percentage of long
amyloid beta protein secreted by familial amyloid beta protein precursor
(beta APP717) mutants.  Science 264:1336--1340.

Takada, S., Z. Luthey-Schulten, and P.G. Wolynes. 1999.  Folding
dynamics with nonadditive forces: a simulation study of a designer
helical protein and a random heteropolymer.
J. Chem. Phys. 110:11616--11629.

Thirumalai, D., D.K. Klimov, and R.I.  Dima. 2002.  Insights into
specific problems in protein folding using simple concepts.
Adv. Chem. Phys. 120:35--76.

Thompson, L.K. 2003.  Unraveling the secrets of Alzheimer's
$\beta$-amyloid fibrils.  Proc. Natl. Acad. Sci. 100:383--385.

Tjernberg, L.O., D.J.E. Callaway, A.  Tjernberg, S. Hahne,
C. Lillieh\"{o}\"{o}k, L. Terenius, J.  Thyberg, and C. Nordstedt. 1999.
A molecular model of Alzheimer amyloid $\beta$-peptide fibril formation.
J. Biol. Chem. 274:12619--12625.

Tjernberg, L.O., A. Tjernberg, N. Bark, Y.  Shi, B.P. Ruzsicska, Z. Bu,
J. Thyberg, and D.J.E. Callaway. 2002.  Assembling amyloid fibrils from
designed structures containing a significant amyloid $\beta$-peptide
fragment. Biochem. J. 366:343--351.

T\"or\"ok, M., S. Milton, R. Kayed, P.  Wu, T. McIntire, C.G. Glabe,
and R. Langen. 2002.  Structural and dynamic features of Alzheimer's
A$\beta$ peptide in amyloid fibrils studied by site-directed spin
labeling. J. Biol. Chem. 277:40810--40815.

Vorobjev, Y.N., J.C. Almagro, and J. Hermans. 1998.  Discrimination
between native and intentionally misfolded conformations of proteins:
ES/IS, a new method for calculating conformational free energy that uses
both dynamics simulations with an explicit solvent and an implicit
solvent continuum model.  Proteins 32:399--413.

Vorobjev Y.N. and J. Hermans. 1999.  ES/IS: estimation of conformational
free energy by combining dynamics simulations with explicit solvent with
an implicit solvent continuum model.  Biophys. Chem. 78:195--205.

Vorobjev, Y.N. and J. Hermans. 2001.  Free energies of protein decoys
provide insight into determinants of protein stability.  Protein
Sci. U.S.A. 10:2498--2506.

Walsh, D.M., A. Lomakin, G.B. Benedek, M.M.  Condron, and
D. Teplow. 1997.  Amyloid $\beta$-protein fibrillogenesis: detection of
a protofibrillar intermediate.  J. Biol. Chem. 272:22364--22372.

Walsh, D.M., D.M. Hartley, Y.  Kusumoto, Y. Fezoui, M.M. Condron,
A. Lomakin, G.B.  Benedek, D.J. Selkoe, and D.B. Teplow. 1999.  Amyloid
$\beta$-protein fibrillogenesis: structure and biological activity of
protofibrillar intermediates.  J. Biol. Chem. 274:25945--25952.

Walsh, D.M., I. Klyubin, J.V. Fadeeva, W.K. Cullen, R.  Anwyl,
M.S. Wolfe, M.J. Rowan, D.J. Selkoe. 2002.  Naturally secreted oligomers
of amyloid b protein potently inhibit hippocampal long-term potentiation
{\it in vivo}. Nature 416:535--539.

Weggen, S., J.L. Ericksen, P. Das, S.A.  Sagi, R. Wang, C.U. Pietrzik,
K.A. Findlay, T.W.  Smith, M.P. Murphy, T. Butler, D. E. Kang,
N. Marquez-Sterling, T.E. Golde, and E.H.  Koo.  2001.  A subset of
NSAIDs lower amyloidogenic A beta 42 independently of cyclooxygenase
activity.  Nature 414:212--216.

Westerman, M.A., D. Cooper-Blacketer, A. Mariash, L. Kotilinek,
T. Kawarabayashi, L.H. Younkin, G.A. Carlson, S.G. Younkin, and
K.H. Ashe.  2002.  The relationship between Ab and memory in the Tg2576
mouse model of Alzheimer's disease. J. Neuroscience 22:1858--1867.

Wolynes, P.G., Z. Luthey-Schulten, and J.N. Onuchic. 1996.  Fast folding
experiments and the topography of protein folding energy landscapes.
Chem. Biol. 3:425--432.

Xia, W., J. Zhang, D. Kholodenko, M. Citron, M.B.  Podlisny,
D.B. Teplow, C. Haass, P. Seubert, E.H. Koo, and D.J. Selkoe. 1995.
Enhanced production and oligomerization of the 42-residue amyloid
$\beta$-protein by Chinese hampster ovary cells stably expressing mutant
presinilins.  J. Biol. Chem. 272:7977--7982.

Yankner, B.A. 1996. Mechanisms of neuronal degeneration in Alzheimer's
disease. Neuron 16:921--932.

Yong, W., A. Lomakin, M.D. Kirkitadze, D.B.  Teplow, S.-H. Chen, and
G.B. Benedek. 2002.  Structure determination of micelle-like
intermediates in amyloid b-protein fibril assembly by using small angle
neutron scattering.  Proc. Natl. Acad. Sci. 99:150--154.

Zagorski, M.G. and C.J. Barrow. 1992.  NMR studies of amyloid
beta-peptides: proton assignments, secondary structure, and mechanism of
an alpha-helix---beta-sheet conversion for a homologous, 28-residue,
N-terminal fragment.  Biochemistry 31:5621--5631.

Zagrovic, B., C.D. Snow, M.R. Shirts, and V.S. Pande. 2002.  Simulation
of folding of a small $\alpha$-helical protein in atomistic detail using
worldwide-distributed computing.  J. Mol. Biol. 323:927--937.

Zhou, Y. and M. Karplus. 1997.  Folding thermodynamics of a
three-helix-bundle protein.
Proc. Natl. Acad. Sci. U.S.A. 94:14429--14432.
}

%\end{thebibliography}
%======================================================================
\vfill
\eject

\newpage
%======================================================================
%\arabic{table}
%----------------------Table I
\begin{table}[h]
\caption{Free energies of monomer conformations.  Comparison of
calculated free energies, G$_{A\beta-40}$ and G$_{A\beta-42}$, and the
corresponding standard deviations, $\sigma_G$, of A$\beta$ monomer
conformations as determined by the NMR experiment (Coles et al., 1998).
The names of different monomer structures follow the ID code name 1BA4
of Brookhaven Protein Database Bank (http://www.rcsb.org/pdb/).  The
free energy unit is kcal/mol. \\}
\label{table_of_monomers}
\begin{tabular}{|l|c|c|c|c|}
\toprule
Monomer     &    G$_{A\beta-40}$  & $\sigma_G$ & G$_{A\beta-42}$   &  $\sigma_G$  \\ 
\colrule                                                                         
										
1BA4-01     &    -1036.58         &  74.97     &  -1026.23         &  73.48       \\
										
1BA4-02     &    -1050.25         &  77.16     &  -1034.13         &  75.12       \\
										
1BA4-03     &    -1045.88         &  78.46     &  -1028.07         &  73.56       \\
										
1BA4-04     &    -1045.93         &  75.66     &  -1032.92         &  78.13       \\
										
1BA4-05     &    -1030.62         &  75.01     &  -1008.66         &  73.68       \\
										
1BA4-06     &     -997.14         &  72.41     &  -1017.85         &  74.81       \\
										
1BA4-07     &    -1043.71         &  75.81     &  -1039.30         &  75.11       \\
										
1BA4-08     &    -1016.94         &  73.36     &  -1027.37         &  75.40       \\
										
1BA4-09     &    -1038.70         &  75.97     &  -1032.68         &  75.13       \\
										
1BA4-10     &    -1052.29         &  78.13     &  -1044.28         &  74.37       \\ 

\botrule
\end{tabular}
\end{table}

%======================================================================
\vfill
\eject

\newpage
%======================================================================
%\arabic{table}
%----------------------Table II
\begin{table}[h]
\caption{Free energies of dimer conformations.  Comparison of calculated
free energies G$_{A\beta-40}$ and G$_{A\beta-42}$, the corresponding
standard deviations $\sigma_G$, and the free energy differences
$\Delta$G$_{A\beta-40}$ and $\Delta$G$_{A\beta-42}$ of A$\beta$(1-42)
and A$\beta$(1-40) dimer conformations.  The free energy unit is
kcal/mol. \\}
\label{table_of_dimers}
\begin{tabular}{|l|c|c|c|c|c|c|}
\toprule
Dimers           & G$_{A\beta-40}$ & $\sigma_G$ & $\Delta$G$_{A\beta-40}$ & G$_{A\beta-42}$ & $\sigma_G$ & $\Delta$G$_{A\beta-42}$ \\ 
\colrule                                                                                                                           
																   
NN-para          &  -1983.51       & 159.91     &  88.10                  &  -1994.72       & 147.09     &   63.58                 \\
																   
NN-anti          &  -2061.09       & 147.16     &  10.52                  &  -2019.14       & 145.00     &   39.15                 \\
																   
NC-para          &  -1935.59       & 138.00     & 136.02                  &  -1937.60       & 137.21     &  120.70                 \\
																   
NC-anti          &  -1999.38       & 149.00     &  72.23                  &  -1982.94       & 142.48     &   75.36                 \\
																	
CC-para          &  -2000.17       & 143.05     &  71.44                  &  -1871.82       & 131.86     &  186.48                 \\
																	   
CC-anti          &  -2043.70       & 147.02     &  27.91                  &  -2022.44       & 144.31     &   35.86                 \\
																		   
nest-para        &  -1964.90       & 139.27     & 106.71                  &  -1989.43       & 143.00     &   68.87                 \\
																		   
nest-anti        &   unstable      &  N/A       &    N/A                  &  -1950.07       & 228.23     &  108.23                 \\
																		   
anti-nest-para   &  -2028.34       & 204.34     &  43.27                  &  -2022.27       & 207.26     &   36.03                 \\
																 
anti-nest-anti   &  -1988.27       & 205.89     &  83.34                  &  -1972.38       & 205.40     &   85.92                 \\

\botrule
\end{tabular}
\end{table}

\vfill
\eject
%======================================================================

\newpage
% ********* FIGURE LEGENDS ****************** 
\section*{FIGURE LEGENDS}

\begin{description}
\item{Fig.~1:} {Conformations of an A$\beta$(1--42) monomer
peptide model as a function of temperature, (a) mostly
$\alpha$-helix conformation at $T=0.100$ with two $\alpha$
helices at residues 12-23 and 29-38, and a hinge at residues
23-28; (b) 3-$\beta$-strand conformation at $T=0.108$, (c)
4-$\beta$-strand conformation at $T=0.115$, and (d)
$\beta$-hairpin conformation at $T=0.120$ characterized by a
$\beta$-turn at residues 23-28.}

\item{Fig.~2:} {Two distributions that give the probability
for an amino acid at a residue number (position in the chain)
to be within a $\beta$-turn.  These simulations are done at
temperature $T = 0.125$, where our model for A$\beta$(1-42)
yields a stable $\beta$-hairpin conformation.  The curve with
open circles corresponds to the altered chain (no glycines)
and the curve with triangles corresponds to the original
A$\beta$(1-42) model with six glycines.  The distributions
are calculated on the basis of 28 (the model with no
glycines) and 38 (the A$\beta$(1-42) model) different
$\beta$-hairpin configurations.  For each $\beta$-hairpin
conformation we use VMD (Humphrey et al., 1996) visualization
package to determine and count all the residues with a
$\beta$-turn.  The probability to be in the $\beta$-turn is
determined as a ratio between the number of conformations, in
which the amino acid is part of a $\beta$-turn, and the total
number of conformations.}

\item{FIG.~3:} {Schematic conformations of an A$\beta$(1--42)
dimer peptide model.  All the conformations are based on a
$\beta$-hairpin conformation with a $\beta$-turn at residues
23-28.  In our model the energies of all these conformations
are approximately the same, however, the probability of the
occurrence varies.}
\end{description}

\vfill
\eject

% ********* FIGURES ****************** 
\newpage
{\bf Fig.~1}
%---------------FIGURE 1
\begin{figure}[h]
\begin{center}
$\begin{array}{c@{\hspace{0.7cm}}c@{\hspace{0.7cm}}c@{\hspace{0.7cm}}c}
\includegraphics*[width=5.0cm]{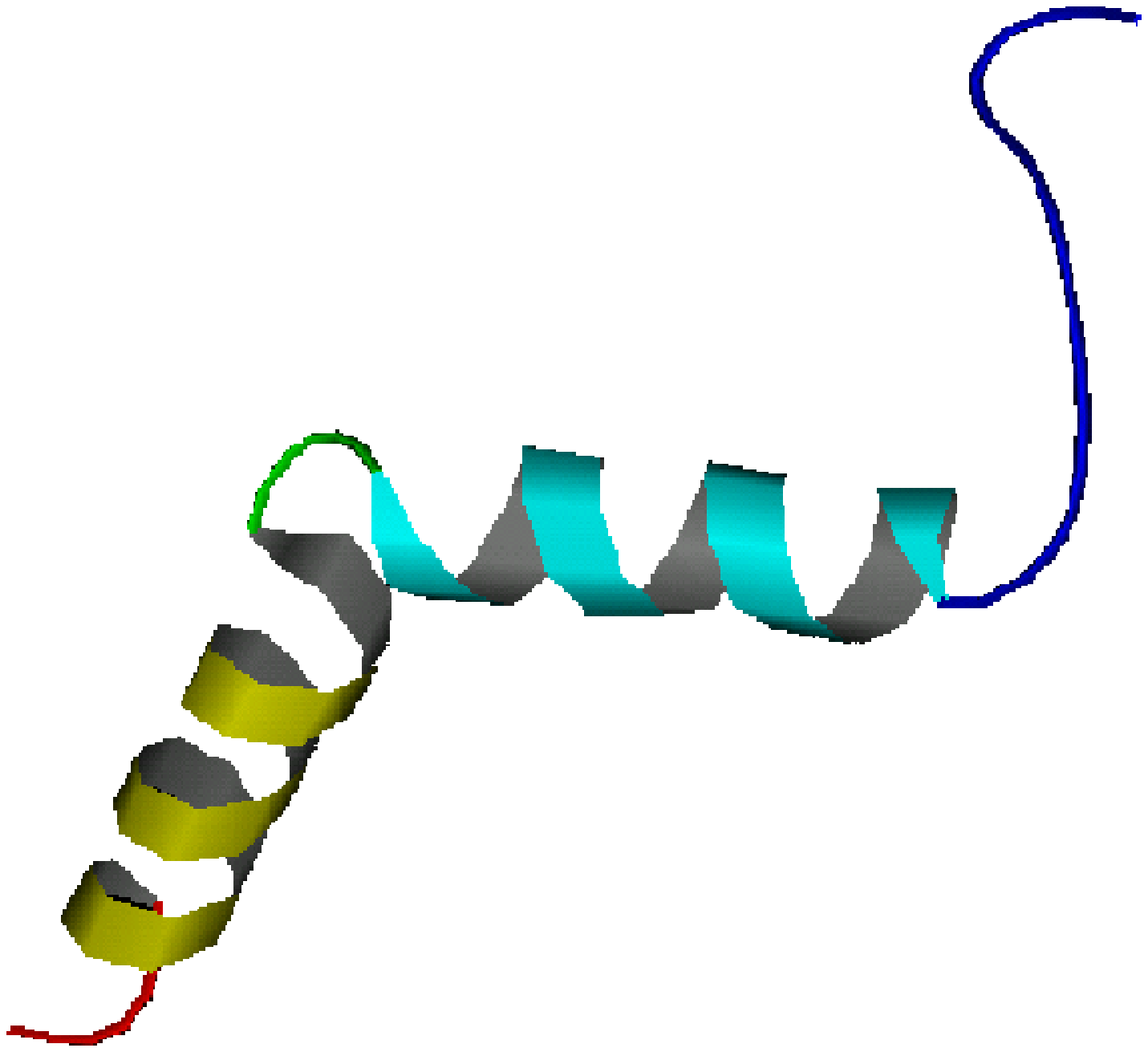}
&
\includegraphics*[width=5.0cm]{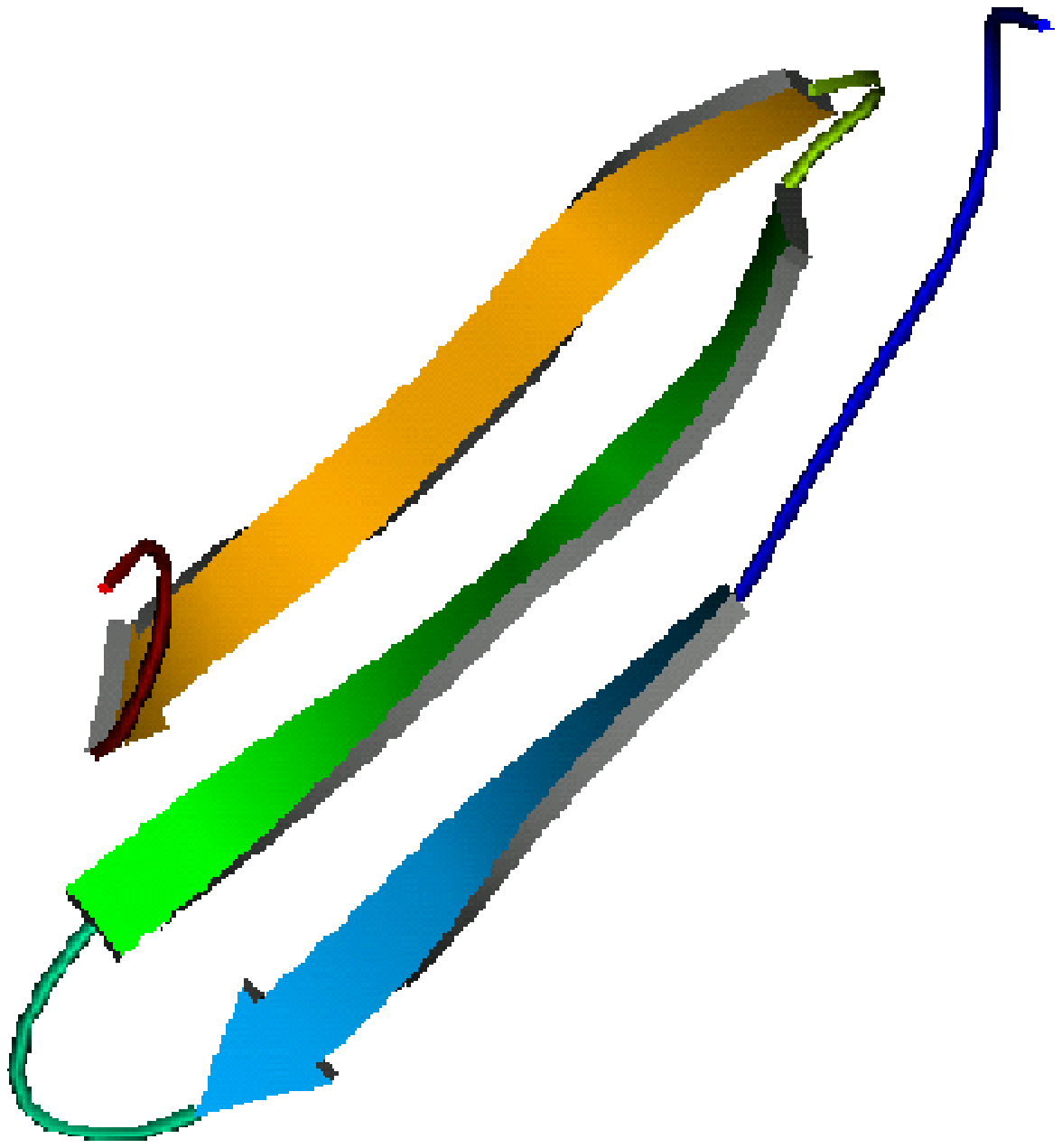}
\\ \text{(a) $T=0.100$} & \text{(b) $T=0.108$}
\end{array}$
$\begin{array}{c@{\hspace{0.7cm}}c@{\hspace{0.7cm}}c@{\hspace{0.7cm}}c@{\hspace{0.7cm}}c}
\includegraphics*[width=5.0cm]{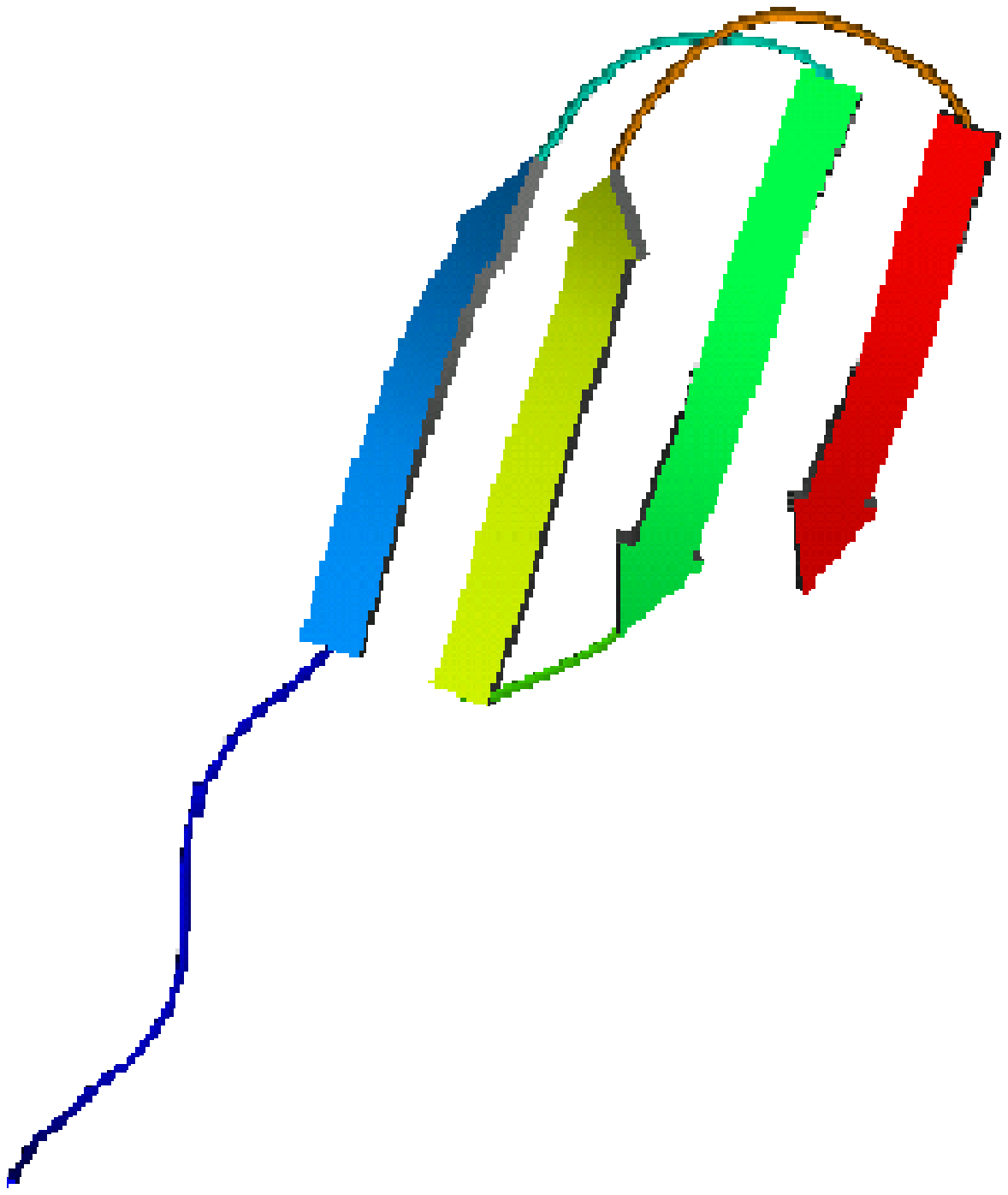}
&
\includegraphics*[width=5.0cm]{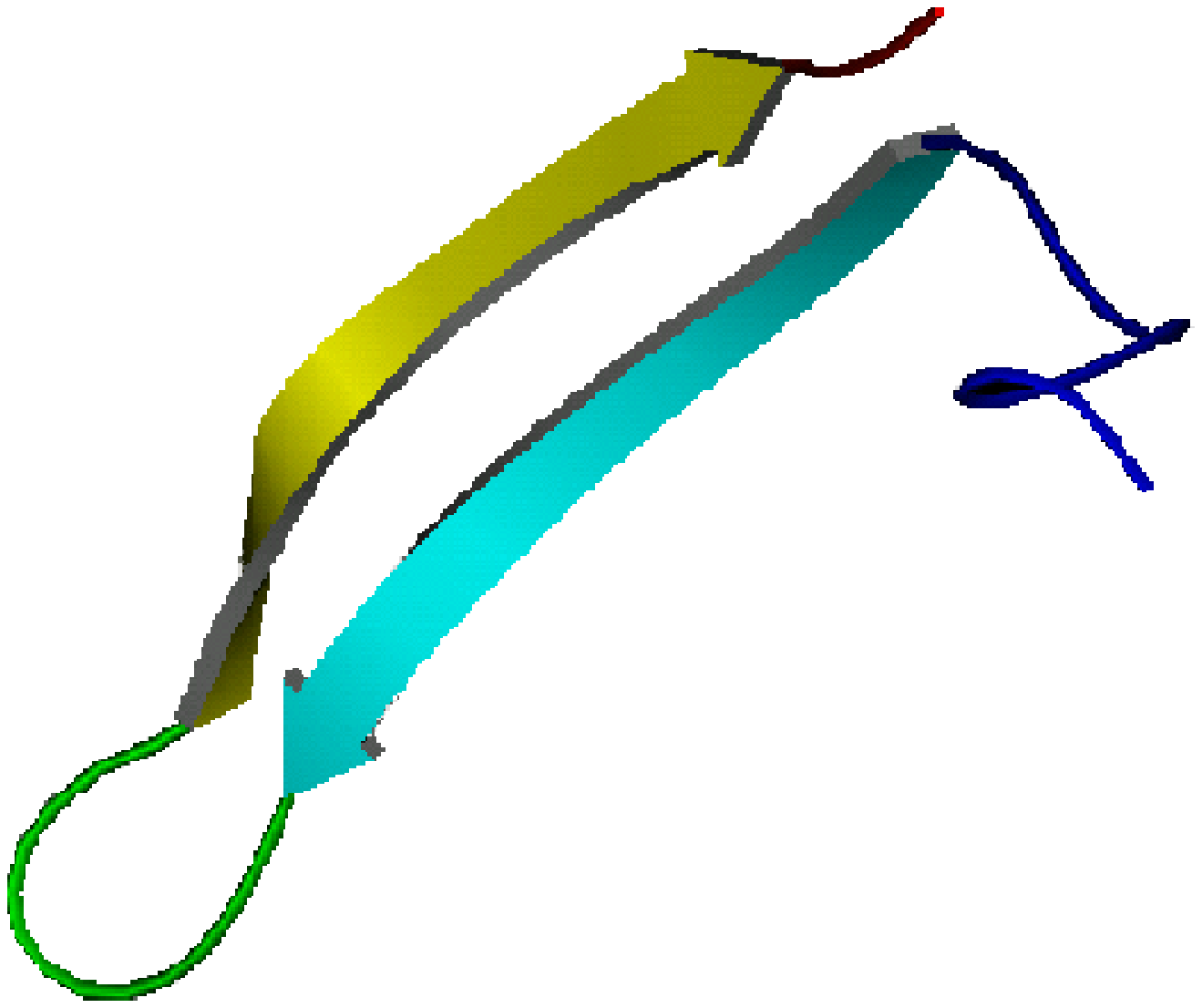}
\\ \text{(c) $T=0.115$} & \text{(d) $T=0.120$}
\end{array}$
\end{center}
%\caption{}
\label{MonomerCharacterization}
\end{figure}
%---------------FIGURE 1
\vfill
\eject

\newpage
{\bf Fig.~2}
%--------------FIGURE 2
\begin{figure}[h]
\begin{center}
$\begin{array}{c@{\hspace{0.7cm}}c@{\hspace{0.7cm}}c}
\includegraphics*[width=10.0cm]{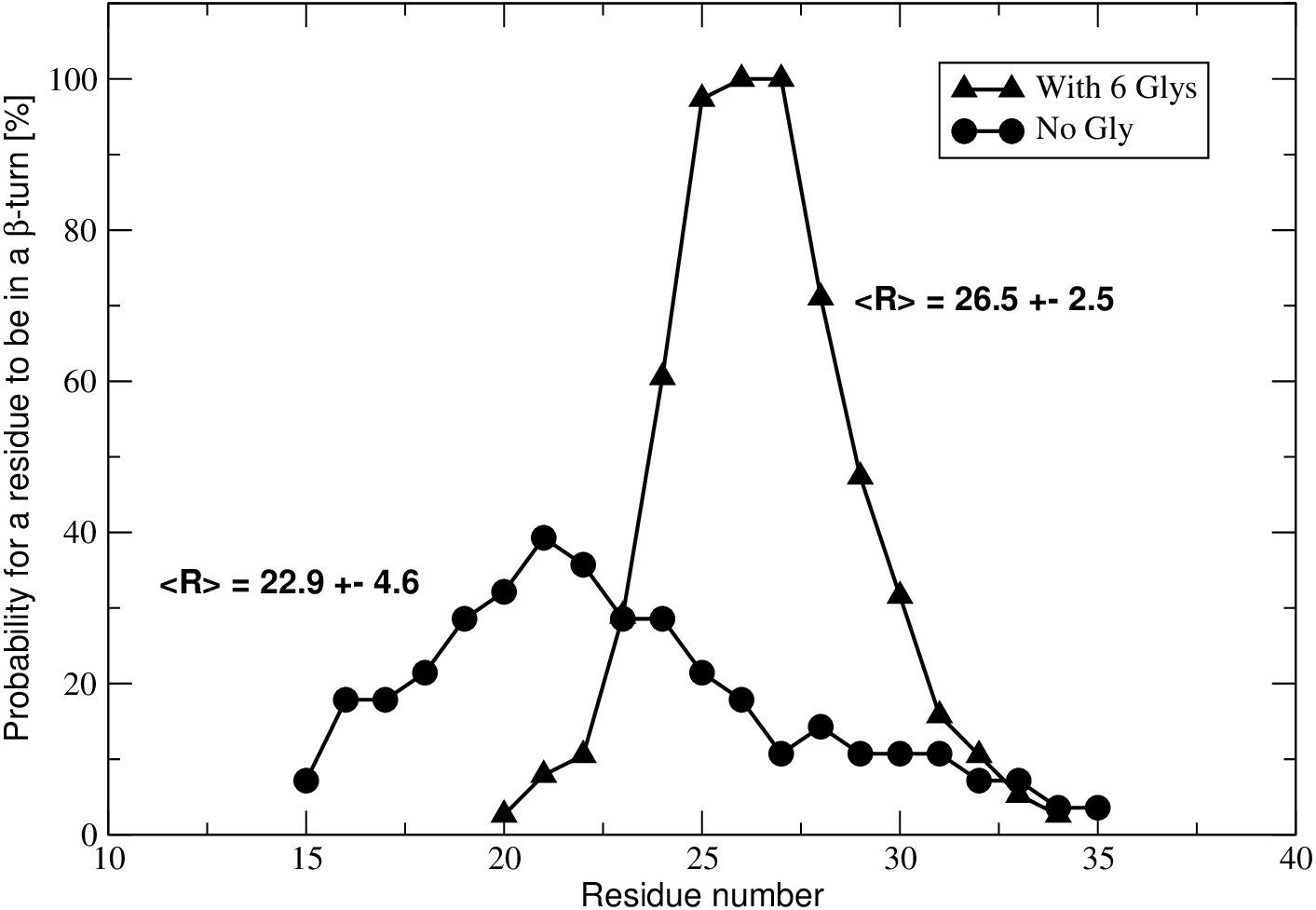}
\\
\end{array}$
\end{center} 
%\caption{}
\label{BetaTurnPosition}
\end{figure}
%--------------FIGURE 2
\vfill
\eject

\newpage
{\bf Fig.~3}
%---------------FIGURE 3
\begin{figure}[h]
\begin{center}
$\begin{array}{c@{\hspace{0.5cm}}c@{\hspace{0.5cm}}c@{\hspace{0.5cm}}c}
\includegraphics*[width=1.5cm]{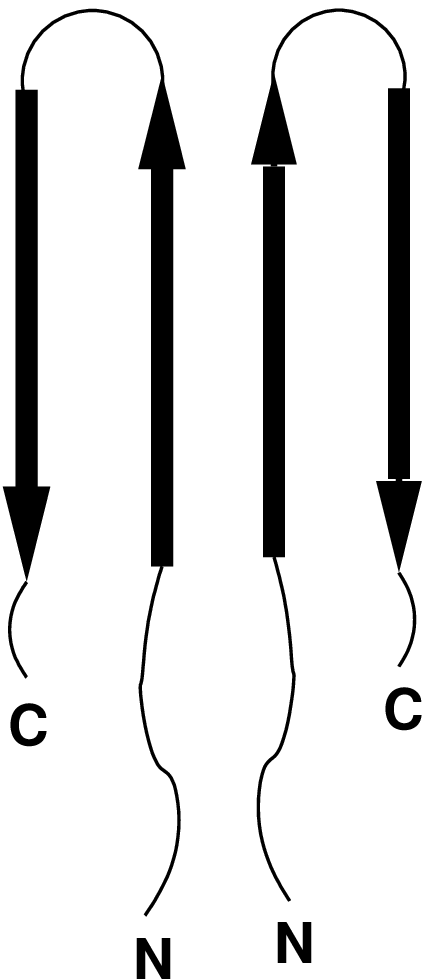}
&
\includegraphics*[width=1.5cm]{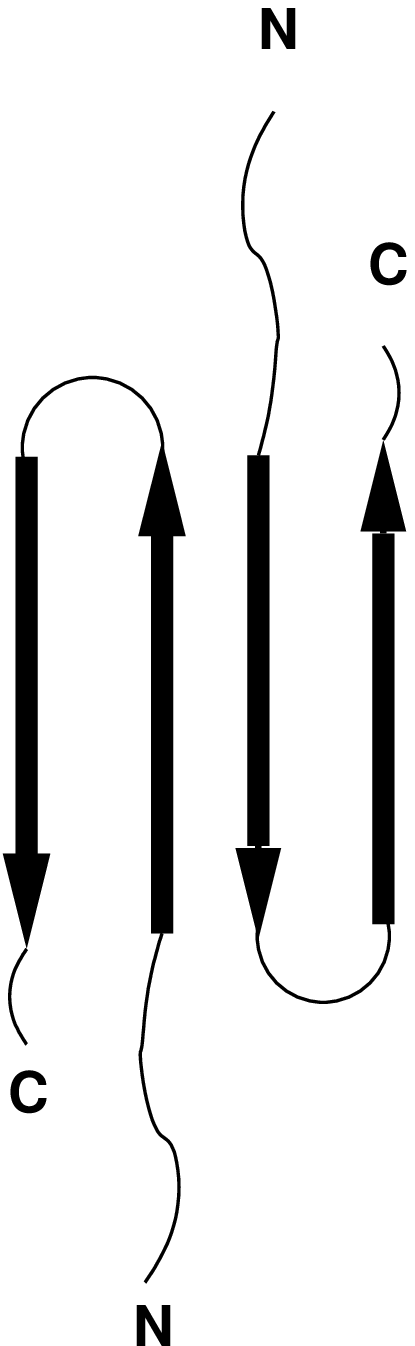}
&
\includegraphics*[width=1.5cm]{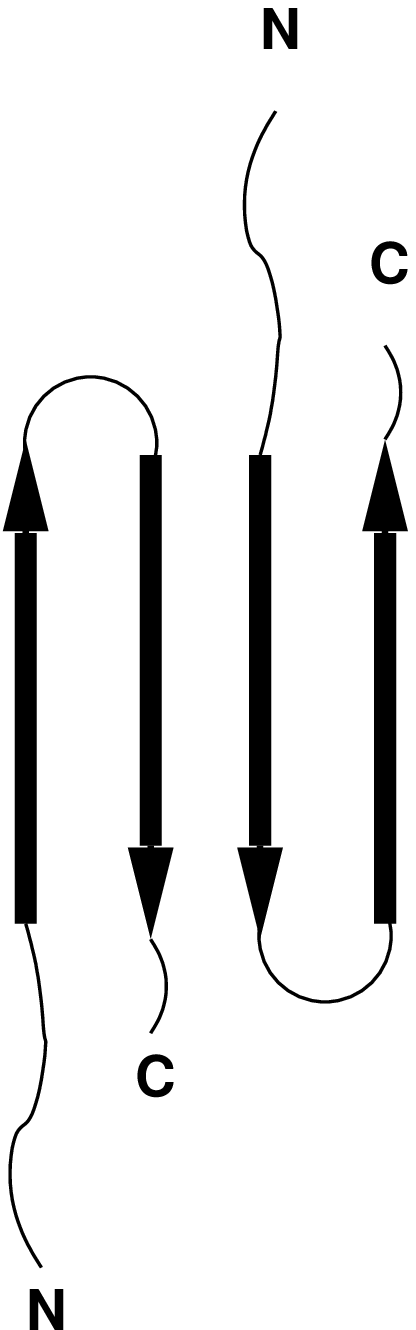}
\\ \text{(a) NN parallel} &  \text{(b) NN antiparallel} & \text{(c) NC parallel} 
\\
\includegraphics*[width=1.5cm]{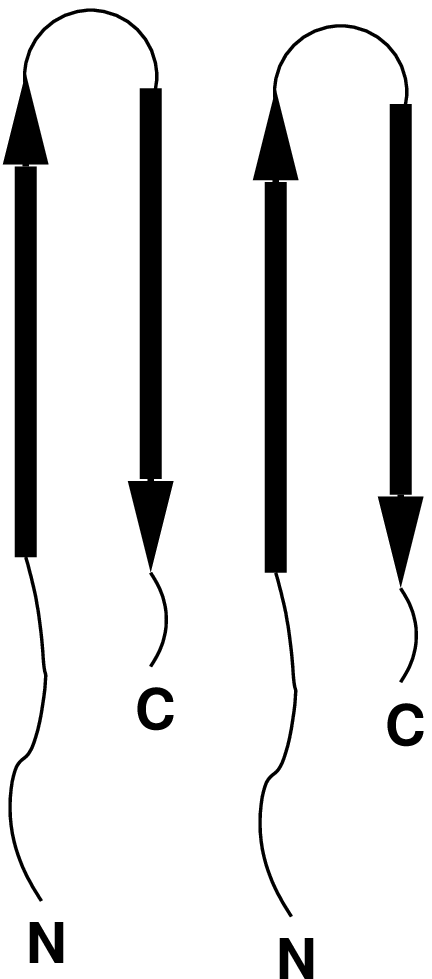}
&
\includegraphics*[width=1.5cm]{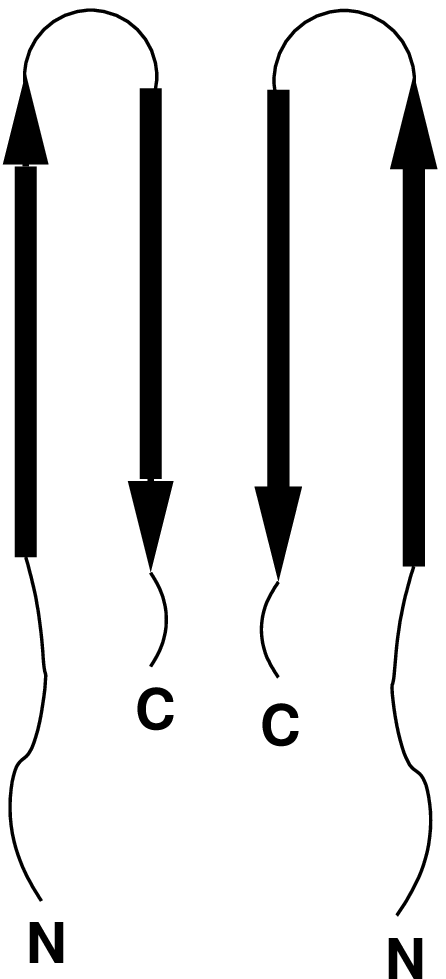}
&
\includegraphics*[width=1.5cm]{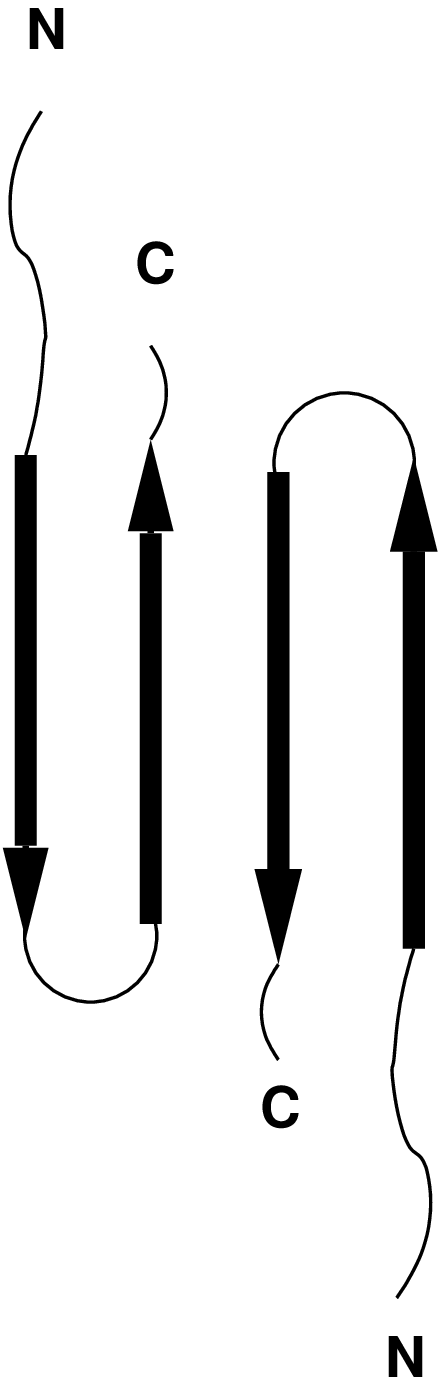}
\\ \text{(d) NC antiparallel} & \text{(e) CC parallel} &  \text{(f) CC antiparallel} 
\\
\includegraphics*[width=1.5cm]{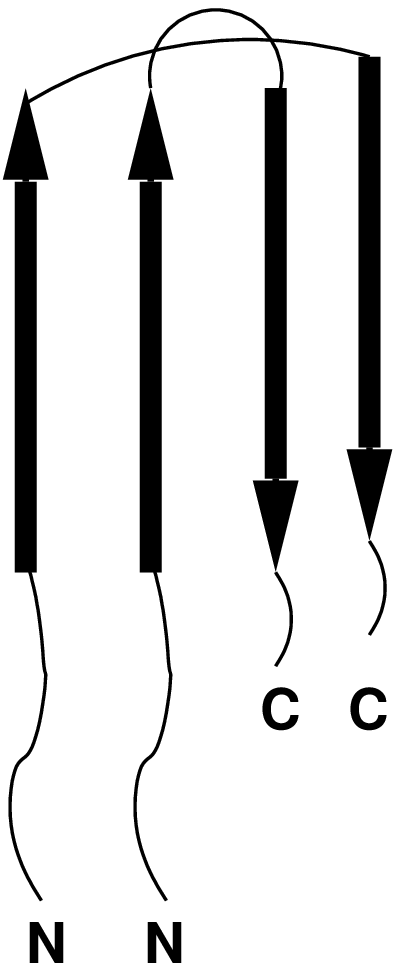}
&
\includegraphics*[width=1.5cm]{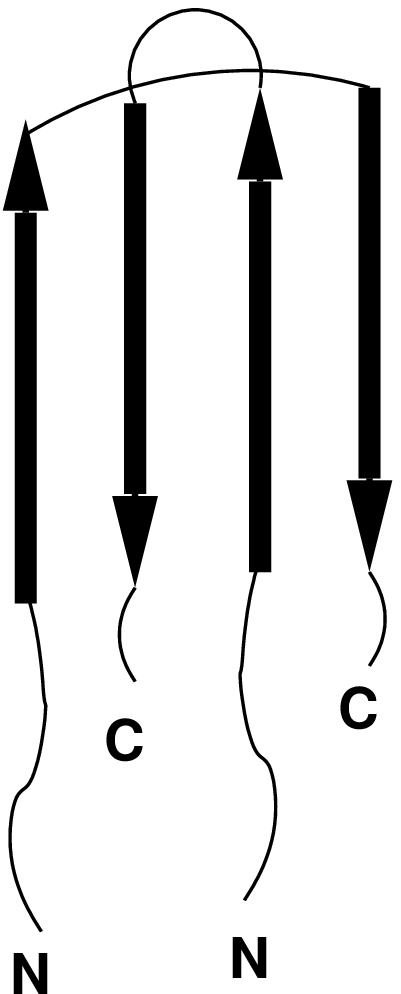}
&
\includegraphics*[width=1.5cm]{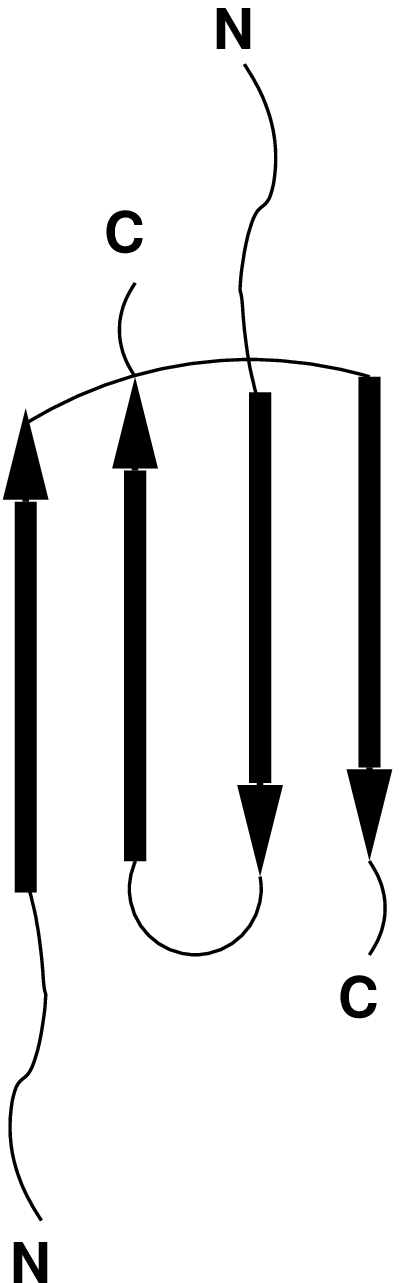}
&
\includegraphics*[width=1.5cm]{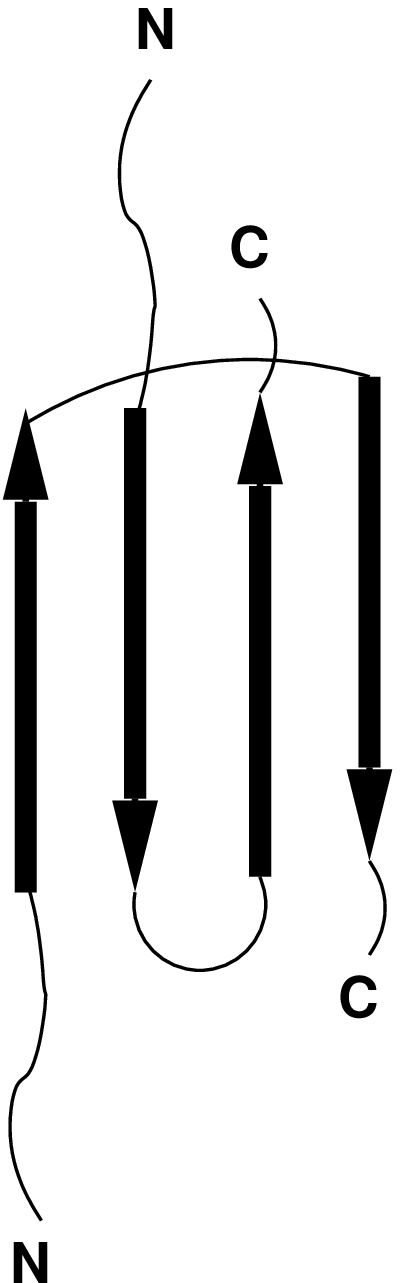}
\\ \text{(g) nested parallel} & \text{(h) nested antiparallel} &  \text{(i) anti-nested parallel} &  \text{(j) anti-nested antiparallel}
\end{array}$
\end{center}
%\caption{}
\label{SchematicDimers}
\end{figure}
%---------------FIGURE 3

%============================================================= 
% *********************************** 
%=============================================================
\vfill 
\eject
\end{document}